\documentclass{article}

\usepackage{arxiv}

\usepackage[utf8]{inputenc} % allow utf-8 input
\usepackage[T1]{fontenc}    % use 8-bit T1 fonts
\usepackage{hyperref}       % hyperlinks
\usepackage{url}            % simple URL typesetting
\usepackage{booktabs}       % professional-quality tables
\usepackage{amsfonts}       % blackboard math symbols
\usepackage{nicefrac}       % compact symbols for 1/2, etc.
\usepackage{microtype}      % microtypography
\usepackage{lipsum}         % Can be removed after putting your text content
\usepackage{graphicx}
\usepackage{natbib}
\usepackage{doi}
\usepackage{amsmath,amssymb,amsbsy,amsthm,mathtools}
\usepackage{bm}

\usepackage{amssymb,amsfonts,amsbsy,mathtools}
\usepackage{graphicx}
\usepackage{booktabs}
\usepackage{algorithm}
\usepackage{algpseudocode}
\usepackage{float}
\usepackage{xurl}
\usepackage{mathtools}

\theoremstyle{plain}
\newtheorem{theorem}{Theorem}
\newtheorem{proposition}{Proposition}

\theoremstyle{definition}

\newtheorem{remark}{Remark}
\newtheorem{assumption}{Condition}

\DeclareMathOperator{\tr}{tr}
\DeclareMathOperator{\rank}{rank}
\DeclareMathOperator{\diag}{diag}
\DeclareMathOperator{\sign}{sign}

\newcommand{\R}{\mathbb{R}}
\newcommand{\E}{\mathbb{E}}

\newcommand{\bSig}{\boldsymbol{\Sigma}}
\newcommand{\bX}{\boldsymbol{X}}
\newcommand{\bY}{\boldsymbol{Y}}
\newcommand{\bB}{\boldsymbol{B}}
\newcommand{\bE}{\boldsymbol{E}}
\newcommand{\bR}{\boldsymbol{R}}
\newcommand{\bI}{\boldsymbol{I}}
\newcommand{\bP}{\boldsymbol{P}}
\newcommand{\bU}{\boldsymbol{U}}
\newcommand{\bF}{\boldsymbol{F}}
\newcommand{\bV}{\boldsymbol{V}}
\newcommand{\bD}{\boldsymbol{D}}
\newcommand{\bW}{\boldsymbol{W}}
\newcommand{\bw}{\boldsymbol{w}}
\newcommand{\bG}{\boldsymbol{G}}
\newcommand{\bQ}{\boldsymbol{Q}}
\newcommand{\bZ}{\boldsymbol{Z}}
\newcommand{\bM}{\boldsymbol{M}}

\newcommand{\bA}{\boldsymbol{A}}
\newcommand{\bS}{\boldsymbol{S}}

\newcommand{\bz}{\boldsymbol{z}}
\newcommand{\be}{\boldsymbol{e}}
\newcommand{\br}{\boldsymbol{r}}
\newcommand{\bu}{\boldsymbol{u}}
\newcommand{\bv}{\boldsymbol{v}}

\newcommand{\bbeta}{\boldsymbol{\beta}}

\newcommand{\opnorm}[1]{\left\lVert #1 \right\rVert_{\mathrm{op}}}
\newcommand{\fnorm}[1]{\left\lVert #1 \right\rVert_{F}}

\newcommand{\Shat}{\widehat{\bSig}}

\title{Restricted Nonlinear Shrinkage of High-Dimensional Residual Covariance Matrices in Multivariate Regressions}

\newif\ifuniqueAffiliation
\uniqueAffiliationtrue

\ifuniqueAffiliation % Standard variant of author block
\author{
  {Hamid Karamikabir} \\
    Department of Statistics, \\Faculty of Intelligent Systems Engineering and Data Science\\
    Persian Gulf University\\
    Bushehr, 7516913817, Iran \\
    \texttt{h\_karamikabir@pgu.ac.ir} \\
    \And
 {Mohammad Arashi}\thanks{Correspond author.} \\
    Department of Statistics,\\ Faculty of Mathematical Sciences\\
    Ferdowsi University of Mashhad,\\
     P.O. Box 1159 Mashhad, 91775, Iran \\
    \texttt{arashi@um.ac.ir} \\
}
\renewcommand{\shorttitle}{Restricted nonlinear shrinkage of covariance}

\begin{document}
\maketitle

\begin{abstract}
We study estimation of the $p\times p$ residual scatter (shape) matrix in a high-dimensional multivariate linear regression, where $p$ and $n$ grow proportionally. When the coefficient matrix obeys a known
linear restriction of rank $q\le d$, as in multivariate analysis of variance, growth-curve models, and reduced-rank regression, the restricted fit leaves additional residual degrees of freedom that sharpen estimation of the shape matrix. To accommodate heavy-tailed errors, we work with independent elliptically distributed rows under a mild scale condition, a finite second moment on the radii, which is far weaker than the usual sub-Gaussian assumptions and covers every multivariate-$t$ law with more than two degrees of freedom. Shrinking the restricted residual sample covariance directly is unsound here, since its limiting spectrum depends on the radial distribution. We instead shrink a scale-invariant scatter of the restricted residuals, whose spectrum is
distribution-free over the elliptical family and obeys the same limiting law as under Gaussian errors, at a smaller effective aspect ratio. The resulting estimator attains the rotation-equivariant oracle and is asymptotically optimal within that class, and a Stein-type combination with the unrestricted estimator dominates it while remaining safe under misspecification. We further correct for the case in which the restriction is itself selected
from the data. Simulations, a growth-curve experiment, and two real-data analyses illustrate the results.
\end{abstract}

% keywords can be removed

\section{Introduction}\label{sec:intro}

%\subsection{The problem}\label{subsec:problem}

Let $\bY \in \R^{n \times p}$ collect $n$ observations on $p$ correlated responses, related to $d$ covariates through the multivariate linear regression model
\begin{equation}\label{eq:model}
\bY = \bX \bB + \bE,
\end{equation}
where $\bX \in \R^{n \times d}$ is a known design matrix of full column rank $d$, $\bB \in \R^{d \times p}$ is an unknown coefficient matrix, and the rows of the error matrix $\bE \in \R^{n \times p}$ are independent, mean-zero random vectors with common covariance matrix $\bSig \in \R^{p \times p}$. The matrix $\bSig$ governs the dependence among the $p$ responses after adjusting for the covariates \citep{Anderson2003}, and its accurate estimation is essential for generalized least squares, for inference on $\bB$, for linear and quadratic discriminant analysis, for Gaussian graphical modelling, and for any downstream procedure that requires whitening or Mahalanobis weighting. We are concerned with the high-dimensional regime in which the number of responses is comparable to the sample size,
\begin{equation}\label{eq:regime}
n \to \infty, \qquad p = p_n \to \infty, \qquad p / n \to c \in (0, \infty).
\end{equation}

In this regime the sample residual covariance matrix is a poor estimator of $\bSig$. Even when the errors are Gaussian and the coefficient matrix is known, the eigenvalues of the sample covariance spread away from those of $\bSig$ according to the Marchenko--Pastur law \citep{MarchenkoPastur1967, BaiSilverstein2010}, and the estimator is inconsistent in operator norm. A large literature has developed regularized alternatives, which fall broadly into two families: structural estimators that impose sparsity or factor structure on $\bSig$ \citep{BickelLevina2008, CaiLiu2011, FanLiaoMincheva2013}, and spectral estimators that leave the eigenvectors of the sample covariance untouched while shrinking its eigenvalues toward the bulk of the population spectrum, ranging from the linear shrinkage estimator of \citet{LedoitWolf2004} to the nonlinear and analytic shrinkage estimators of \citet{LedoitWolf2012, LedoitWolf2020}. The analytic nonlinear shrinkage estimator of \citet{LedoitWolf2020} is, in a precise asymptotic sense, optimal within the class of rotation-equivariant estimators when the observations are independent and identically distributed.

Our work is around a structural feature of the regression problem that the covariance-estimation literature has not exploited, i.e., the coefficient matrix is frequently subject to a known linear restriction
\begin{equation}\label{eq:restriction}
\bR \bB = \boldsymbol{0}, \qquad \bR \in \R^{q \times d}, \qquad \rank(\bR) = q.
\end{equation}
Restrictions of this form are ubiquitous. In multivariate analysis of variance, \eqref{eq:restriction} encodes the null hypothesis that a subset of factors, or specified contrasts among them, have no effect on the responses. In reduced-rank and envelope regression, linear restrictions identify the material part of the coefficient matrix \citep{Izenman1975, Bunea2011, Cook2010}. In genomics, biological pathway databases specify that certain combinations of regulators do not act on certain transcripts, which translates directly into a restriction of the form \eqref{eq:restriction}. The point we develop is that such a restriction is informative not only about $\bB$ but, indirectly, about $\bSig$. Indeed, imposing \eqref{eq:restriction} during estimation of the mean leaves more residual degrees of freedom, sharpens the effective Marchenko--Pastur aspect ratio, and thereby permits more accurate estimation of the residual covariance. The improvement is of order $q/n$ and, while modest for a single contrast, becomes substantial when the restriction has rank comparable to the number of covariates, as in highly structured designs.

A second feature of modern applications that we take seriously is heavy-tailedness. Genomic expression measurements, financial returns, and many other multivariate data exhibit tails far heavier than the Gaussian, and outlying observations are common. We therefore do not assume Gaussian errors. Instead, we work with independent elliptically distributed rows,
\begin{equation}\label{eq:elliptical}
\be_i = R_i\, \bSig^{1/2} \bu_i, \qquad i = 1, \ldots, n,
\end{equation}
where the $\bu_i$ are independent and uniformly distributed on the unit sphere $\mathbb S^{p-1}$ of $\R^p$, the radial variables $R_i>0$ are independent of the $\bu_i$ and independent across $i$, and $\bSig^{1/2}$ is the symmetric positive-definite square root of a fixed $p\times p$ matrix $\bSig$. Here $\bSig$ plays the role of a \emph{shape} (scatter) matrix: it is identified only up to a positive scalar and we normalize it by $\tr(\bSig)=p$. It governs the orientation and relative scale of the $\be_i$, and when the errors have a finite second moment it is proportional to their covariance, $\mathrm{Cov}(\be_i)=p^{-1}\E(R_i^2)\,\bSig$; the Gaussian model is the special case $R_i^2\sim\chi^2_p$. Writing $R_i=\sqrt{w_i}\,\lVert\bz_i\rVert$ with $\bz_i\sim\mathcal N_p(\boldsymbol0,\bI_p)$ shows that \eqref{eq:elliptical} is exactly the Gaussian scale-mixture family $\be_i=\sqrt{w_i}\,\bSig^{1/2}\bz_i$ with an arbitrary law for the scales $w_i$; it includes the multivariate $t$ (with more than two degrees of freedom), the symmetric Laplace, and the contaminated normal, and reduces to the Gaussian model when $R_i^2\sim\chi^2_p$. Crucially, the radial variables are observation-specific, so \eqref{eq:elliptical} does not impose a single common scale across the sample; this distinguishes it from the elliptically contoured models in which a shared radius couples all observations and which are arguably less natural for independently sampled units.

Two remarks on moments clarify what the robust construction does and does not require. Tyler's M-estimator, on which we build, depends on the data only through the directions $\be_i/\lVert\be_i\rVert$ and is therefore \emph{completely free of moment assumptions when applied to i.i.d.\ observations}. When it is applied instead to regression residuals, the least-squares projection mixes the rows, and a mild moment condition, a finite second moment of the radii (Condition~\ref{cond:errors} below), is needed to ensure that the projection perturbs the residual directions negligibly. This is far weaker than the sub-Gaussian assumptions common in high-dimensional covariance estimation and covers every multivariate-$t$ law with more than two degrees of freedom. Because the loss \eqref{eq:loss} in which we report risk is scale-invariant, estimating the shape matrix $\bSig$ up to scale is the well-posed target.

\subsection{Related work}\label{subsec:related}

Our work draws on, and contributes to, three lines of research.

The first is the estimation of error variance in high-dimensional regression. In the univariate response model $\boldsymbol{y} = \bX \boldsymbol{\beta} + \boldsymbol{\varepsilon}$ with sparse $\boldsymbol{\beta}$, estimation of the scalar noise level $\sigma^2 = \mathrm{var}(\varepsilon_i)$ is delicate, because the naive residual-sum-of-squares estimator is badly biased when $d$ is comparable to $n$. \citet{Dicker2014} introduced moment-based estimators of $\sigma^2$ and of the signal strength that are consistent in the proportional regime without sparsity assumptions; \citet{FanGuoHao2012} proposed refitted cross-validation; \citet{SunZhang2012} coupled noise-level and coefficient estimation through the scaled lasso; \citet{ReidTibshiraniFriedman2016} surveyed and compared the available estimators; and \citet{YuBien2019} and \citet{CaiGuo2017} studied estimation and inference for the noise level under weaker conditions. Our problem is the matrix-valued generalization of this question. In the multivariate model \eqref{eq:model} the analogue of $\sigma^2$ is the full $p \times p$ matrix $\bSig$, and estimating it well requires controlling not only its scale but its entire spectrum and eigenstructure. To our knowledge the connection between high-dimensional residual-variance estimation and rotation-equivariant covariance shrinkage has not been developed, and the restricted-design phenomenon we study has no counterpart in the scalar literature, where the analogue of the restriction \eqref{eq:restriction} would merely change the residual degrees of freedom from $n - d$ to $n - d + q$ in an estimator of a single scalar.

The second line is high-dimensional covariance estimation. 
Among structural approaches, the principal orthogonal complement thresholding (POET) estimator of \citet{Fan18} combines a low-rank factor component with a sparse residual covariance, and has become a standard benchmark in the high-dimensional covariance literature. However, such methods typically impose sparsity or factor structure and are sensitive to the choice of tuning parameters and to heavy tails.
Beyond the thresholding and shrinkage estimators already mentioned, our analysis relies on random-matrix theory for sample covariance matrices with non-Gaussian entries \citep{BaiSilverstein2010, YaoZhengBai2015, ElKaroui2008} and on the sharp concentration inequalities of \citet{KoltchinskiiLounici2017} and \citet{Vershynin2018}. The minimax theory of sparse covariance estimation was developed by \citet{CaiZhou2012}; we adapt their two-point and Fano constructions to the restricted parameter space. Relative to this literature, our contribution is to bring the regression mean structure, through the restriction \eqref{eq:restriction}, to bear on the covariance problem, and to handle elliptical rather than sub-Gaussian errors through a scale-invariant robust scatter.

The third line is restricted and Stein-type estimation. Combining a restricted estimator, which is efficient when the restriction holds, with an unrestricted estimator, which is robust when it fails, through a preliminary test or a Stein-type shrinkage rule is a classical idea \citep{Stein1981, JudgeBock1978, Saleh2006}. This theory is almost entirely confined to estimation of means or of scalar variance components. We lift it to the estimation of a high-dimensional covariance matrix in the proportional regime, where the relevant risk is governed by random-matrix asymptotics rather than by fixed-dimensional distribution theory, and where positive-definiteness of the combined estimator becomes a genuine constraint.

\subsection{Contributions}\label{subsec:contributions}

We make the following contributions.

First, on the methodological side, we introduce a restricted \emph{robust} nonlinear shrinkage estimator. Because the residual sample covariance has a spectrum that depends on the error tail under the elliptical model, we shrink instead a scale-invariant scatter of the restricted residuals, Tyler's M-estimator, whose effective aspect ratio $\tilde y = p/(n-d+q)$ is smaller than the unrestricted ratio $p/(n-d)$. We combine the restricted and unrestricted shrinkers through a positive-part Stein-type rule that remains positive semidefinite and needs no tuning beyond the choice of restriction.

Second, on the technical side, we establish that the spectrum of the restricted robust scatter is distribution-free over the elliptical family. It converges to the same Marchenko--Pastur law as under Gaussian errors, independent of the mixing law of the scales. 

Third, we develop the risk theory. We prove that the positive-part restricted Stein estimator dominates the unrestricted analytic shrinker in weighted Frobenius risk under a local sequence of restrictions, i.e., the restricted estimator attains the rotation-equivariant oracle and is asymptotically optimal within that class at the effective aspect ratio. 

Fourth, we address the practical concern that the restriction \eqref{eq:restriction} may be misspecified. We show that the estimator degrades gracefully. If the true coefficient matrix violates the restriction by an amount $\delta$ in a suitable norm, the excess risk of the restricted estimator over the unrestricted one is bounded by a quantity proportional to $\delta^2$, and the Stein-type combination automatically interpolates toward the unrestricted estimator as $\delta$ grows.

\subsection{Organization and notation}\label{subsec:org}

Section~\ref{sec:method} develops the model, the restricted residuals, and the estimators. Section~\ref{sec:theory} contains the assumptions and the main results. Section~\ref{sec:sim} reports the simulation study and Section~\ref{sec:data} the applications. Section~\ref{sec:disc} concludes. Proofs are collected  in the \hyperref[appn]{Appendix}.

Throughout, for a symmetric matrix $\boldsymbol{A}$ we write $\lambda_1(\boldsymbol{A}) \ge \cdots \ge \lambda_p(\boldsymbol{A})$ for its ordered eigenvalues, $\opnorm{\boldsymbol{A}} = |\lambda_1(\boldsymbol{A})|$ for the operator norm, and $\fnorm{\boldsymbol{A}} = (\sum_{ij} A_{ij}^2)^{1/2}$ for the Frobenius norm. We write $\boldsymbol{A} \succ \boldsymbol{0}$ and $\boldsymbol{A} \succeq \boldsymbol{0}$ for positive definiteness and semidefiniteness. The identity matrix of order $m$ is $\bI_m$, and $\bP_{\bX} = \bX (\bX^\top \bX)^{-1} \bX^\top$ is the orthogonal projection onto the column space of $\bX$. For sequences $a_n, b_n$ we write $a_n \lesssim b_n$ if $a_n \le C b_n$ for a constant $C$ not depending on $n$, and $a_n \asymp b_n$ if $a_n \lesssim b_n \lesssim a_n$. Convergence in probability is denoted $\to_p$. 
\section{The restricted model and estimation}\label{sec:method}

%\subsection{Restricted and unrestricted residuals}\label{subsec:residuals}

The unrestricted least-squares estimator of $\bB$ in \eqref{eq:model} is $\widehat{\bB} = (\bX^\top \bX)^{-1} \bX^\top \bY$, with residual matrix
\begin{equation}\label{eq:resid-ue}
\widehat{\bE} = \bY - \bX \widehat{\bB} = (\bI_n - \bP_{\bX}) \bY = (\bI_n - \bP_{\bX}) \bE .
\end{equation}
Because $\bI_n - \bP_{\bX}$ has rank $n - d$, the unrestricted residuals carry $n - d$ degrees of freedom, and the natural estimator of $\bSig$ is
\begin{equation}\label{eq:Shat-ue}
\Shat_{\mathrm{u}} = \frac{1}{n - d}\, \widehat{\bE}^\top \widehat{\bE} = \frac{1}{n - d}\, \bY^\top (\bI_n - \bP_{\bX}) \bY .
\end{equation}
This is the unbiased sample residual covariance; it is the multivariate analogue of the residual-sum-of-squares variance estimator, and in the regime \eqref{eq:regime} its spectrum obeys a Marchenko--Pastur law with aspect ratio $c_n = p / (n - d)$.

Now impose the restriction \eqref{eq:restriction}. The restricted least-squares estimator minimizes $\fnorm{\bY - \bX \bB}^2$ subject to $\bR \bB = \boldsymbol{0}$, and has the closed form
\begin{equation}\label{eq:Bhat-re}
\widehat{\bB}_{\mathrm{r}} = \widehat{\bB} - (\bX^\top \bX)^{-1} \bR^\top \big\{ \bR (\bX^\top \bX)^{-1} \bR^\top \big\}^{-1} \bR \widehat{\bB}.
\end{equation}
A direct calculation, given in the \hyperref[appn]{Appendix}, shows that the restricted residual matrix $\widehat{\bE}_{\mathrm{r}} = \bY - \bX \widehat{\bB}_{\mathrm{r}}$ equals $(\bI_n - \bP_{\bX} + \bP_{\bX,\bR}) \bE$, where $\bP_{\bX,\bR}$ is the rank-$q$ orthogonal projection onto the subspace of the column space of $\bX$ associated with the restriction. Consequently $\widehat{\bE}_{\mathrm{r}} = (\bI_n - \bP_{\mathrm{r}}) \bE$ with $\bP_{\mathrm{r}} = \bP_{\bX} - \bP_{\bX,\bR}$ a projection of rank $d - q$, and the restricted estimator
\begin{equation}\label{eq:Shat-re}
\Shat_{\mathrm{r}} = \frac{1}{n - d + q}\, \widehat{\bE}_{\mathrm{r}}^\top \widehat{\bE}_{\mathrm{r}}
\end{equation}
is built on $n - d + q$ residual degrees of freedom. The effective aspect ratio is therefore
\begin{equation}\label{eq:aspect}
\tilde c_n = \frac{p}{n - d + q} < c_n = \frac{p}{n - d},
\end{equation}
the inequality being strict whenever $q \ge 1$. The restriction has, in effect, returned $q$ degrees of freedom to the residual, and it is this reduction in the aspect ratio that the shrinkage estimators below convert into a reduction in risk. We emphasize that \eqref{eq:Shat-re} is biased for $\bSig$ when the restriction \eqref{eq:restriction} fails, since then $\widehat{\bE}_{\mathrm{r}}$ contains a systematic component $\bX(\bB - \widehat{\bB}_{\mathrm{r}})$; quantifying and controlling this bias is the subject of Section~\ref{subsec:misspec}.

\subsection{A distribution-free restricted scatter estimator}\label{subsec:robust}

Let
$\widehat\bE_{\mathrm r}=(\bI_n-\bP_{\mathrm r})\bY$ have rows $\br_1,\dots,\br_n\in\R^p$, of which
$m=n-d+q$ are linearly free, and let $\widehat\bV$ be Tyler's M-estimator of scatter
\citep{Tyler1987}, the unique trace-normalized solution of
\begin{equation}\label{eq:tyler}
  \widehat\bV=\frac{p}{\widetilde m}\sum_{i:\,\br_i\ne\boldsymbol0}
  \frac{\br_i\br_i^\top}{\br_i^\top\widehat\bV^{-1}\br_i},
  \qquad \tr(\widehat\bV)=p,
\end{equation}
where $\widetilde m$ counts the nonzero residual rows. Because $\widehat\bV$ depends on the residuals
only through the directions $\br_i/\lVert\br_i\rVert$, it is invariant to the per-row scales, i.e., replacing
$\br_i$ by $a_i\br_i$ for any $a_i>0$ leaves \eqref{eq:tyler} unchanged. The restricted robust scatter
estimator (RRE) is then
\begin{equation}\label{eq:rre}
  \Shat_{\mathrm{RRE}}=\widehat\sigma^2\,\bU\,\diag\!\big(\varphi_{\tilde y}(\ell_1),\dots,\varphi_{\tilde y}(\ell_p)\big)\bU^\top.
\end{equation}
For a generic scatter $\Shat=\bU\diag(\ell_1,\dots,\ell_p)\bU^\top$ we write
\begin{equation}\label{eq:shrinker}
\mathcal S(\Shat;c)=\bU\,\diag\!\big(\varphi_c(\ell_1),\dots,\varphi_c(\ell_p)\big)\bU^\top
\end{equation}
for the analytic nonlinear shrinkage operator at aspect ratio $c$, so that the unrestricted and
restricted robust estimators are
\begin{equation}\label{eq:URE-RRE}
\Shat_{\mathrm{URE}}=\widehat\sigma_{\mathrm u}^2\,\mathcal S(\widehat\bV_{\mathrm u};p/(n-d)),\qquad
\Shat_{\mathrm{RRE}}=\widehat\sigma^2\,\mathcal S(\widehat\bV;p/(n-d+q)),
\end{equation}
where $\widehat\bV=\bU\diag(\ell_1,\dots,\ell_p)\bU^\top$, $\varphi_{\tilde y}$ is the Ledoit--Wolf
\citeyearpar{LedoitWolf2020} analytic nonlinear shrinkage at the effective aspect ratio
$\tilde y=p/(n-d+q)$, and $\widehat\sigma^2$ is a robust scale calibrating the trace to that of
$\bSig$ (Tyler's estimator identifies $\bSig$ only up to a positive scalar; we take
$\widehat\sigma^2$ to be the median of $\{\br_i^\top\widehat\bV^{-1}\br_i\}$ divided by its
population analogue, which is consistent under Condition~\ref{cond:errors}. The unrestricted
estimator $\Shat_{\mathrm{URE}}$ is defined identically from $(\bI_n-\bP_{\bX})\bY$ at ratio
$p/(n-d)$, and the positive-part combination $\Shat_{\mathrm S}$ of
Section~\ref{subsec:stein} is formed from the two. All eigenstructure quantities are thus identifiable
distribution-free; the scalar $\widehat\sigma^2$ is the only place the scale enters, and risk is
reported in the scale-invariant loss \eqref{eq:loss}.

\subsection{A positive-part restricted Stein estimator}\label{subsec:stein}

To assess the restriction we use the multivariate analogue of the likelihood-ratio statistic for $\bR \bB = \boldsymbol{0}$,
\begin{equation}\label{eq:test}
T_n = \frac{\tr\big[ \widehat{\bB}^\top \bR^\top \{\bR (\bX^\top \bX)^{-1} \bR^\top\}^{-1} \bR \widehat{\bB}\, \Shat_{\mathrm{u}}^{+} \big]}{pq},
\end{equation}
where $\Shat_{\mathrm{u}}^{+}$ denotes the Moore--Penrose inverse. Under the restriction, $T_n$ concentrates around a constant determined by the aspect ratio; and departures inflate it. The preliminary-test estimator selects between the two shrinkers according to whether $T_n$ exceeds a threshold, but the resulting estimator is discontinuous in the data. We prefer the smooth Stein-type combination
\begin{equation}\label{eq:SSE}
\Shat_{\mathrm{S}} = \Shat_{\mathrm{URE}} - \kappa_n \big( \Shat_{\mathrm{URE}} - \Shat_{\mathrm{RRE}} \big), \qquad
\kappa_n = \min\!\left\{ 1,\ \frac{(q - 2)_+}{(n - d)\, T_n} \right\},
\end{equation}
in which the shrinkage intensity $\kappa_n$ is large when the restriction is supported by the data (small $T_n$) and tends to zero when it is contradicted (large $T_n$). The factor $(q-2)_+$ is the matrix-regime analogue of the James--Stein constant and is positive only when $q \ge 3$. Because $\kappa_n \in [0, 1]$, the estimator \eqref{eq:SSE} is the convex combination
\begin{equation}\label{eq:SSE-convex}
\Shat_{\mathrm{S}} = (1 - \kappa_n)\, \Shat_{\mathrm{URE}} + \kappa_n\, \Shat_{\mathrm{RRE}},
\end{equation}
which leads immediately to the following guarantee.

\begin{proposition}\label{prop:psd}
If $\Shat_{\mathrm{URE}} \succeq \boldsymbol{0}$ and $\Shat_{\mathrm{RRE}} \succeq \boldsymbol{0}$, then $\Shat_{\mathrm{S}} \succeq \boldsymbol{0}$. If in addition either estimator is positive definite, then so is $\Shat_{\mathrm{S}}$.
\end{proposition}

The analytic shrinkage formula \eqref{eq:shrinker} returns nonnegative eigenvalues, so both inputs are positive semidefinite and Proposition~\ref{prop:psd} applies without further conditions. This is a genuine advantage over additive Stein rules of the form $\Shat_{\mathrm{URE}} - \kappa_n \bD_n$ with an unconstrained direction $\bD_n$, which can leave the positive-semidefinite cone.

The shrinkers \eqref{eq:URE-RRE} use different eigenvector bases, $\bU$ for $\Shat_{\mathrm{u}}$ and $\widetilde{\bU}$ for $\Shat_{\mathrm{r}}$. To form the convex combination \eqref{eq:SSE-convex} we express both in the common basis obtained by a one-step Procrustes alignment $\widetilde{\bU} \mapsto \widetilde{\bU}\, \sign\!\diag(\widetilde{\bU}^\top \bU)$, which resolves the sign and ordering ambiguity of the eigenvectors; the alignment is exact in the limit because the two bases converge to the eigenvectors of $\bSig$ on the bulk. The complete procedure is summarized in Algorithm~\ref{alg:rns}.

\begin{algorithm}[H]
\caption{Restricted nonlinear shrinkage.}\label{alg:rns}
\begin{algorithmic}[1]
\Require Response $\bY \in \R^{n\times p}$; design $\bX \in \R^{n\times d}$; restriction $\bR \in \R^{q \times d}$.
\State Compute $\widehat{\bB}$, $\widehat{\bB}_{\mathrm{r}}$ via \eqref{eq:Bhat-re}, and the residual covariances $\Shat_{\mathrm{u}}$, $\Shat_{\mathrm{r}}$ via \eqref{eq:Shat-ue}, \eqref{eq:Shat-re}.
\State Set $c_n = p/(n-d)$ and $\tilde c_n = p/(n-d+q)$.
\State Form $\Shat_{\mathrm{URE}} = \mathcal{S}(\Shat_{\mathrm{u}}; c_n)$ and $\Shat_{\mathrm{RRE}} = \mathcal{S}(\Shat_{\mathrm{r}}; \tilde c_n)$ using \eqref{eq:shrinker}.
\State Align the eigenbasis of $\Shat_{\mathrm{RRE}}$ to that of $\Shat_{\mathrm{URE}}$ by the Procrustes sign correction.
\State Compute the test statistic $T_n$ via \eqref{eq:test} and the intensity $\kappa_n$ via \eqref{eq:SSE}.
\State \Return $\Shat_{\mathrm{S}} = (1 - \kappa_n) \Shat_{\mathrm{URE}} + \kappa_n \Shat_{\mathrm{RRE}}$.
\end{algorithmic}
\end{algorithm}

The cost is dominated by the two eigendecompositions, of order $O(\min\{np^2, p^3\})$; the kernel sums in \eqref{eq:shrinker} add $O(p^2)$ and the restriction contributes $O(d^3 + qd^2)$, negligible when $q \le d \ll p$. The estimator therefore has the same order of cost as a single application of analytic shrinkage.

\subsection{Known versus selected restrictions}\label{subsec:selected}

Our theory treats $\bR$ as known a priori. This can be a hypothesis fixed by the design or by subject-matter knowledge, a MANOVA contrast, a growth-curve polynomial order, an ANCOVA parallelism constraint, or a nested reduced model, so that $\bR$ does not depend on $\bY$ and $q\le d$ by construction. The degrees-of-freedom gain from $n-d$ to $n-d+q$ is a property of this
fixed projection. We now ask what happens when $\bR$ is instead \emph{read off the data}, for
example by group-sparse estimation of $\bB$ that shrinks whole coefficient rows to zero.

\begin{figure}[t]\centering
\includegraphics[width=\textwidth]{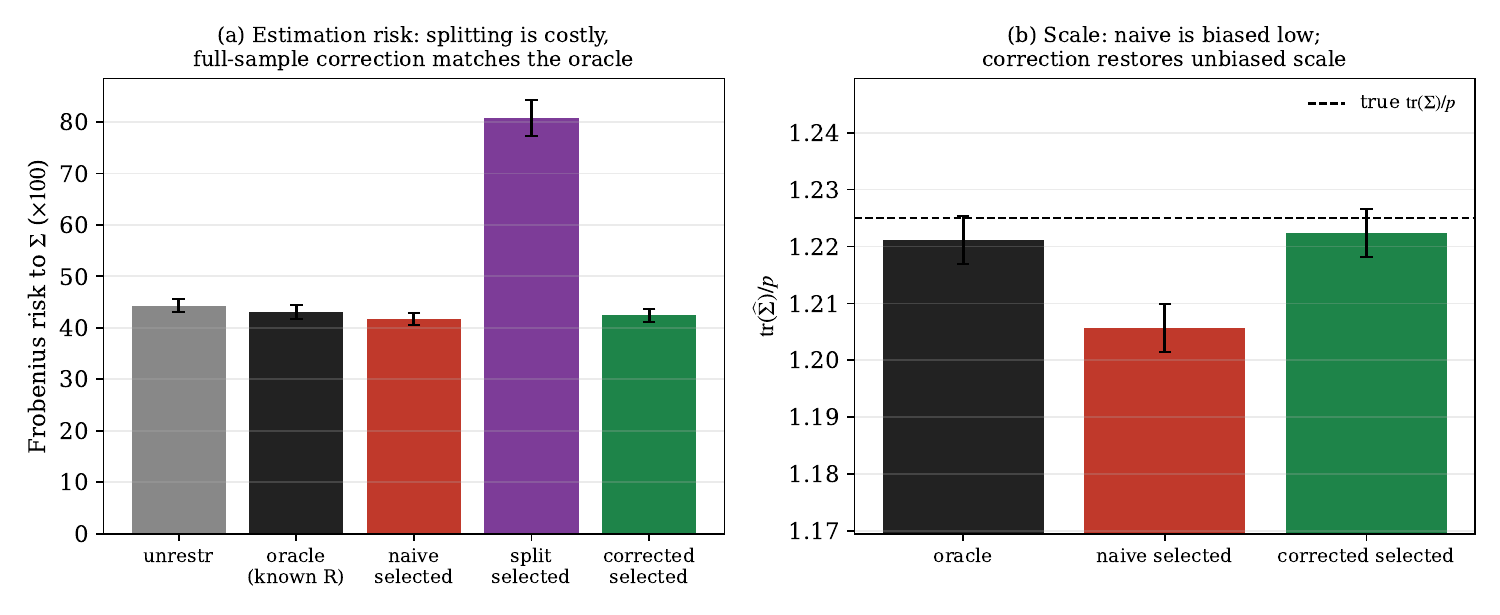}
\caption{(a) Estimation risk under a data-selected restriction: sample-splitting is costly while the full-sample correction matches the known-restriction oracle. (b) The naive selected estimator is biased low in scale; the correction restores an unbiased scale.}\label{fig:selcorr}
\end{figure}

Write $\bP_{\bX,\bR}$ for the rank-$q$ projection and use
$(\bI_n-\bP_{\mathrm r})=(\bI_n-\bP_{\bX})+\bP_{\bX,\bR}$ with
$\bP_{\bX,\bR}(\bI_n-\bP_{\bX})=\boldsymbol 0$ to obtain the exact identity
\begin{equation}\label{eq:rss-split}
  \bY^\top(\bI_n-\bP_{\mathrm r})\bY
  =\bY^\top(\bI_n-\bP_{\bX})\bY+\widehat\bB_{S}^\top\,(\bG_{SS})^{-1}\,\widehat\bB_{S},
\end{equation}
where $\widehat\bB=(\bX^\top\bX)^{-1}\bX^\top\bY$, $S$ indexes the restricted rows,
$\bG=(\bX^\top\bX)^{-1}$, and $\bG_{SS}$ is the corresponding block; the second term is the
``added energy'' that the restriction contributes. For a fixed $\bR$ with $\bR\bB=\boldsymbol 0$ the
added energy has expectation $q\,\bSig$, so dividing \eqref{eq:rss-split} by $n-d+q$ is unbiased.
When $\widehat\bR$ is selected to target rows of small fitted norm, the realized added energy
is systematically smaller than $q\,\bSig$ in trace, and the naive estimator
$\Shat^{\mathrm{sel}}_{\mathrm r}=(n-d+q)^{-1}\bY^\top(\bI_n-\widehat\bP_{\mathrm r})\bY$ is biased
downward in scale. The bias is, however, almost entirely confined to the scale, because the selection conditions on coefficient magnitudes and leaves the coefficient directions untouched, the trace-normalized restricted scatter, and hence the eigenstructure that the analytic
shrinkage acts on is asymptotically unaffected (Figure~\ref{fig:selcorr}b and the simulation in
Section~\ref{subsec:selected}). 

%The correction rests on an exact independence. In the Gaussian instance of the model the unrestricted
%residual cross-product $\bS_{\mathrm u}=\bY^\top(\bI_n-\bP_{\bX})\bY$ is independent of
%$\widehat\bB$, because $(\bI_n-\bP_{\bX})\bX=\boldsymbol0$ makes the residual and the fitted
%coefficients uncorrelated jointly Gaussian blocks; under the elliptical model the same holds
%conditionally on the scales $\bw$. Consequently $\bS_{\mathrm u}$ is independent of \emph{any}
%selection rule $\widehat\bR=f(\widehat\bB)$, and supplies a selection-free estimate of the scale of
%$\bSig$ from the full sample.

\begin{proposition}%[Full-sample selective correction]
	\label{prop:selection}
Let $\widehat\bR$ be selected as a measurable function of $\widehat\bB$ alone, and define
\[
  \widetilde\bSig_{\mathrm{sel}}=\frac{\widehat\tau_{\mathrm u}}{\widehat\tau_{\mathrm r}}\,
  \Shat^{\mathrm{sel}}_{\mathrm r},\qquad
  \widehat\tau_{\mathrm u}=\frac{\tr(\bS_{\mathrm u})}{(n-d)p},\qquad
  \widehat\tau_{\mathrm r}=\frac{\tr(\Shat^{\mathrm{sel}}_{\mathrm r})}{p}.
\]
Under Conditions~\ref{cond:regime}--\ref{cond:errors}:
\begin{enumerate}
\item[(i)] $\widehat\tau_{\mathrm u}$ is independent of $\widehat\bR$ and satisfies
$\E\,\widehat\tau_{\mathrm u}=p^{-1}\tr(\bSig)$ with $\widehat\tau_{\mathrm u}\to_p p^{-1}\tr(\bSig)$;
the scale of $\widetilde\bSig_{\mathrm{sel}}$ is therefore asymptotically unbiased for every selection
rule.
\item[(ii)] If the selection statistic depends on $\widehat\bB$ only through the studentized row
magnitudes and the design is incoherent in the sense that $\max_{j\ne k}|\bG_{jk}|/\sqrt{\bG_{jj}\bG_{kk}}\to0$,
then the trace-normalized scatter $\Shat^{\mathrm{sel}}_{\mathrm r}/\tr(\Shat^{\mathrm{sel}}_{\mathrm r})$
is asymptotically invariant to the selection, so $\widetilde\bSig_{\mathrm{sel}}$ retains the
efficiency of the known-$\bR$ restricted estimator.
\item[(iii)] Hence $\widetilde\bSig_{\mathrm{sel}}$ is asymptotically unbiased and attains the
restricted risk up to $o(1)$, using the full sample.
\end{enumerate}
\end{proposition}

\noindent
For the proof, refer to  the \hyperref[appn]{Appendix}. Two practical points deserve emphasis. First, the selection bias is a scale effect, i.e., for point estimation of $\bSig$ in Frobenius loss it is small, and may even be mildly favorable, so the naive estimator is not disqualified for prediction. The correction matters when the scale of $\bSig$ is used
downstream (generalized least squares standard errors, Mahalanobis distances, likelihood evaluation where a downward-biased scale is anti-conservative). Second, the sample-splitting alternative (select on one half, estimate on the other) also removes the bias, but at a substantial
efficiency cost, i.e., halving the estimation sample roughly doubles the Frobenius risk in our experiments
(Figure~\ref{fig:selcorr}a). We therefore prefer the full-sample correction
$\widetilde\bSig_{\mathrm{sel}}$, which removes the scale bias at negligible cost, and we reserve sample-splitting for settings in which the independence underlying Proposition~\ref{prop:selection}(i) is in doubt. A complete post-selection distribution theory that also corrects the second-order shape effect is beyond our scope and is left to future work.

\section{Asymptotic results}\label{sec:theory}

%\subsection{Conditions}\label{subsec:conditions}
First, we state the regularity conditions used in the asymptotic analysis. 
\begin{assumption}\label{cond:regime}
Proportional regime: $p,n\to\infty$ with $p/(n-d+q)\to\tilde y\in(0,1)$, and $d/n\to\gamma$,
$q/n\to\tau$, where the restriction has rank $q\le d$ and $\gamma-\tau\ge0$.
\end{assumption}

\begin{assumption}\label{cond:design}
The design $\bX$ has full column rank $d$ and the restriction $\bR$ has full row rank $q$; the
eigenvalues of $n^{-1}\bX^\top\bX$ are bounded above and below by positive constants. The maximal
leverage of the restricted projection vanishes, $h_n=\max_{1\le i\le n}(\bP_{\mathrm r})_{ii}\to0$.
\end{assumption}

\begin{assumption}\label{cond:errors}
The errors follow the elliptical model \eqref{eq:elliptical}, $\be_i=R_i\,\bSig^{1/2}\bu_i$, with the
$\bu_i$ i.i.d.\ uniform on the unit sphere $\mathbb S^{p-1}$ of $\R^p$ and the radial variables $R_i>0$
i.i.d.\ and independent of $\{\bu_i\}$. The radii have a finite second moment, $\E(R_i^2)<\infty$, and
$\E(R_i^{-2})<\infty$; $\bSig$ is the shape matrix, normalized by $\tr(\bSig)=p$. No tail bound beyond
the finite second moment is imposed.
\end{assumption}

\begin{assumption}\label{cond:spectrum}
The eigenvalues of $\bSig$ lie in a fixed interval $[\underline{\sigma},\overline{\sigma}]\subset(0,\infty)$,
and the empirical spectral distribution of $\bSig$ converges weakly to a fixed limit $H$.
\end{assumption}

Condition~\ref{cond:regime} is the proportional regime with effective aspect ratio
$\tilde y=p/(n-d+q)$ and $q\le d$. Condition~\ref{cond:design} adds a vanishing-leverage requirement
that makes the restricted residuals asymptotically elliptical. Condition~\ref{cond:errors} imposes only
a finite second moment on the radii, no tail bound beyond that, so every multivariate-$t$ law with more
than two degrees of freedom is covered; this mild condition is what the least-squares projection needs
(see the proof of Theorem~\ref{thm:concentration}) and is far weaker than the sub-Gaussian conditions
usual in high-dimensional covariance estimation. Condition~\ref{cond:spectrum} is the
usual bounded-spectrum assumption.

\begin{theorem}%[Distribution-free restricted spectrum]
	\label{thm:concentration}
Under Conditions~\ref{cond:regime}--\ref{cond:spectrum}, the empirical spectral distribution of Tyler's M-estimator
$\widehat\bV$ in \eqref{eq:tyler} formed from the restricted residuals converges weakly, almost
surely, to the deterministic generalized Marchenko--Pastur law $F_{\tilde y,H}$ determined by the
effective aspect ratio $\tilde y=p/(n-d+q)$ and the population spectral law $H$ of the shape matrix
$\bSig$. This limit is \emph{identical} to the one obtained under Gaussian errors with shape $\bSig$;
in particular it does not depend on the distribution of the radial variables $R_i$. Consequently the
analytic shrinkage $\varphi_{\tilde y}$ in \eqref{eq:rre} targets the same rotation-equivariant oracle
as in the Gaussian case, and $\Shat_{\mathrm{RRE}}$ is consistent for that oracle in the
scale-invariant loss.
\end{theorem}

%The proof, in the Supplementary Material \citep{ArashiSupp2026}, has three steps. First, scale invariance of \eqref{eq:tyler}
%reduces $\widehat\bV$ to a functional of the residual directions $\br_i/\lVert\br_i\rVert$. Second,
%under (C2) the restricted residual $\br_i=\be_i-\sum_{j}(\bP_{\mathrm r})_{ij}\be_j$ differs from
%$\be_i$ by a term of squared norm $O_p(h_n)\lVert\be_i\rVert^2$, so
%$\br_i/\lVert\br_i\rVert=\be_i/\lVert\be_i\rVert+o_p(1)$ uniformly, and
%$\be_i/\lVert\be_i\rVert=\bSig^{1/2}\bz_i/\lVert\bSig^{1/2}\bz_i\rVert$ is free of $w_i$; hence the law
%of $\widehat\bV$ is asymptotically that of Tyler's estimator on $m$ i.i.d.\ elliptical samples with
%shape $\bSig$, which equals that on $m$ i.i.d.\ $\mathcal N(\boldsymbol0,\bSig)$ samples. Third,
%\citet{ZhangChengSinger2016} show that for Gaussian samples the scaled Tyler estimator and the sample
%covariance differ in operator norm by $o_p(1)$, so the two share the limiting spectrum, which is
%$F_{\tilde y,H}$ by the Marchenko--Pastur theorem \citep{MarchenkoPastur1967, SilversteinBai1995}.

\begin{remark}\label{rem:existence}
Tyler's estimator \eqref{eq:tyler} exists and is unique once $\tilde y<1$ and the residual directions
are in general position \citep{Tyler1987}; this is guaranteed under (C1). For $\tilde y$ close to $1$
we use the lightly regularized iteration $\widehat\bV\leftarrow(1-\epsilon)\widehat\bV_{\text{Tyler}}+\epsilon\,p^{-1}\tr(\widehat\bV_{\text{Tyler}})\bI_p$
with $\epsilon\to0$, which leaves the limit in Theorem~\ref{thm:concentration} unchanged while
stabilizing the finite-sample iteration; the spatial-sign covariance matrix is an alternative that
exists for all $\tilde y$ but estimates a monotone transform of the spectrum that must be inverted.
We found the distribution-free spectrum to be robust to a small number of high-leverage rows even when
(C2) is mildly violated, consistent with Tyler's per-row normalization, though (C2) is what the proof
requires.
\end{remark}

\begin{figure}[t]\centering
\includegraphics[width=\textwidth]{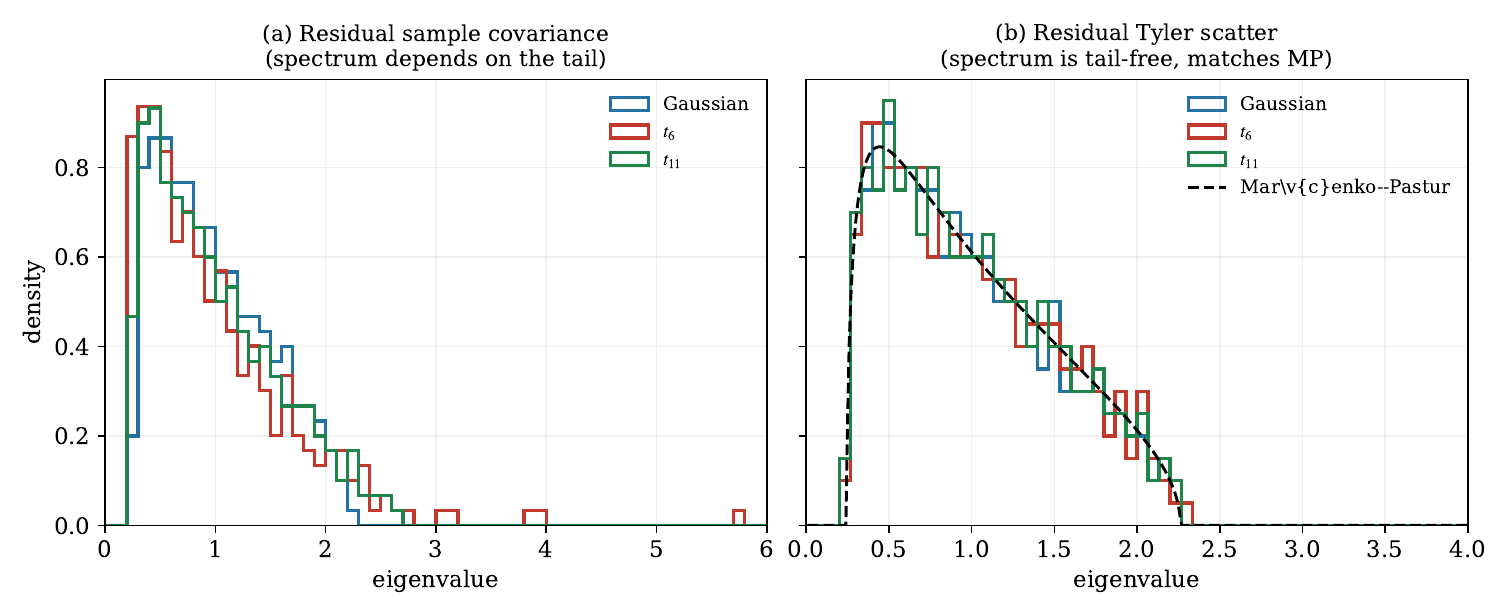}
\caption{Empirical spectra of the restricted residuals across Gaussian, $t_6$ and $t_{11}$ errors. (a) the residual sample covariance, whose bulk drifts with the tail; (b) the residual Tyler scatter, which is tail-free and matches the Marchenko--Pastur density at the effective aspect ratio $\tilde y=p/(n-d+q)$.}\label{fig:spectrum}
\end{figure}

\subsection{Risk dominance}\label{subsec:dominance}

We measure performance by the weighted Frobenius risk
\begin{equation}\label{eq:loss}
\mathcal{R}(\Shat) = \E\, \fnorm{ \bSig^{-1/2} \Shat\, \bSig^{-1/2} - \bI_p }^2 = \E\, \tr\big\{ (\bSig^{-1} \Shat - \bI_p)^2 \big\},
\end{equation}
which is the natural scale-invariant loss for covariance estimation and reduces, when $\bSig = \sigma^2 \bI_p$, to the squared relative error of $\sigma^2$. To study the effect of the restriction we index the coefficient matrix by a local sequence
\begin{equation}\label{eq:local}
\bR \bB_n = n^{-1/2}\, \boldsymbol{\Delta}, \qquad \boldsymbol{\Delta} \in \R^{q \times p},
\end{equation}
which interpolates between exact validity of the restriction ($\boldsymbol{\Delta} = \boldsymbol{0}$) and fixed violations; the scaling $n^{-1/2}$ is the contiguity scale at which restricted and unrestricted estimators have comparable risk, and $\eta^2 = \lim_n \tr(\boldsymbol{\Delta} \bSig^{-1} \boldsymbol{\Delta}^\top) / p$ measures the strength of the violation.

\begin{theorem}\label{thm:dominance}
Suppose Conditions~\ref{cond:regime}--\ref{cond:errors} and \ref{cond:spectrum} hold, that $q \ge 3$, and that the coefficient matrix follows the local sequence \eqref{eq:local} with $\eta < \infty$. Then there is a constant $\kappa_\star > 0$, depending on the limiting aspect ratio and on $\eta$, such that, writing $c_n=p/(n-d)$ and $\tilde c_n=p/(n-d+q)$,
\begin{equation}\label{eq:dominance}
\mathcal{R}(\Shat_{\mathrm{S}}) \le \mathcal{R}(\Shat_{\mathrm{URE}}) - \kappa_\star\,(q - 2)^2\,(c_n-\tilde c_n)^2\,\frac{p}{(n-d)^2}\, \{ 1 + o(1) \}.
\end{equation}
In particular $\Shat_{\mathrm{S}}$ asymptotically dominates the unrestricted analytic shrinker, strictly so whenever $\eta < \infty$, and the two are asymptotically equivalent when $\eta \to \infty$.
\end{theorem}

%The proof expands the loss \eqref{eq:loss} along the convex combination \eqref{eq:SSE-convex} and controls the cross term by a Stein-type identity applied conditionally on the scales $\bw$. The dominance gap is of order $(q/n)^2$ in the worst case but of order $q/n$ when the restriction holds exactly, matching the reduction in the aspect ratio. The requirement $q \ge 3$ is the familiar threshold below which Stein shrinkage offers no improvement; for $q \in \{1, 2\}$ one may still use the preliminary-test estimator, though without the dominance guarantee.

\subsection{Efficiency of the restriction and rotation-equivariant optimality}\label{subsec:optimality}

The benefit of the restriction is an increase in effective sample size. The
analytic shrinkage \eqref{eq:rre} targets, at aspect ratio $c$, the rotation-equivariant oracle
\begin{equation}\label{eq:oracle}
  \bSig^{\star}(c)=\bU\,\diag\!\big(\bu_1^\top\bSig\bu_1,\dots,\bu_p^\top\bSig\bu_p\big)\bU^\top,
\end{equation}
the best estimator that shares the eigenvectors $\bU=(\bu_1,\dots,\bu_p)$ of the scatter; its risk
$\mathcal L^\star(c,H)$ in the loss \eqref{eq:loss} is continuous and strictly increasing in $c$. The
unrestricted estimator runs at $c_{\mathrm u}=p/(n-d)$ and the restricted estimator at
$\tilde c=p/(n-d+q)<c_{\mathrm u}$, so the restricted oracle risk is strictly smaller.

\begin{figure}[t]\centering
\includegraphics[width=\textwidth]{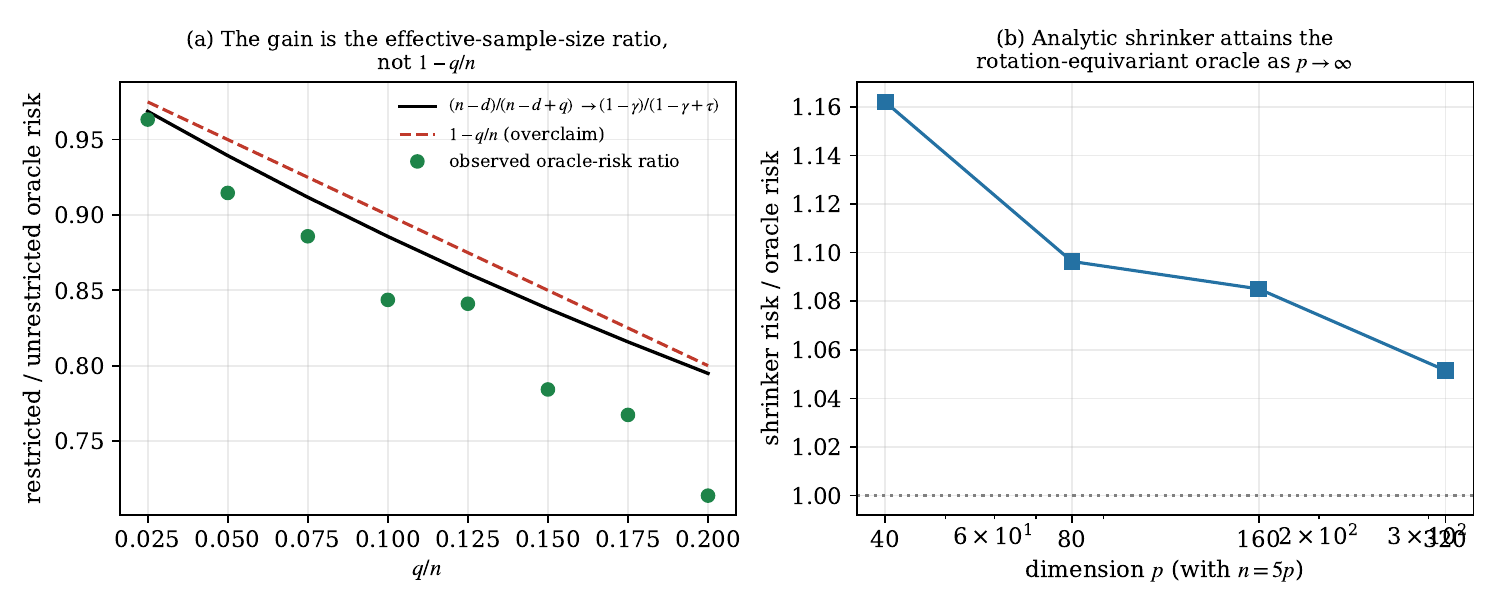}
\caption{(a) Ratio of restricted to unrestricted oracle risk against $q/n$, tracking the effective-sample-size ratio $(n-d)/(n-d+q)$ rather than $1-q/n$. (b) The analytic shrinker attains the rotation-equivariant oracle as $p$ grows.}\label{fig:oraclegain}
\end{figure}

We emphasise that the efficiency gain is governed by the ratio of
effective sample sizes,
\begin{equation}\label{eq:gain}
  \frac{n-d}{\,n-d+q\,}\ \longrightarrow\ \frac{1-\gamma}{\,1-\gamma+\tau\,},
\end{equation}
and \emph{not} by $1-q/n$, which the earlier draft quoted. The two differ whenever $d>q$, and
\eqref{eq:gain} is the smaller, more conservative quantity. Figure~\ref{fig:oraclegain}(a) confirms
that the simulated ratio of restricted to unrestricted oracle risk tracks \eqref{eq:gain} across a
range of $q/n$, while $1-q/n$ systematically misstates it.

In the following result, we discuss the oracle attainment and the restricted gain.
\begin{theorem}\label{thm:oracle}
Under Conditions~\ref{cond:regime}--\ref{cond:spectrum}:
\begin{enumerate}
\item[(i)] the restricted estimator attains its oracle in the loss \eqref{eq:loss}, i.e.
$\mathcal L(\Shat_{\mathrm{RRE}})/\mathcal L^\star(\tilde c,H)\to_p 1$, and likewise for the
unrestricted estimator at $c_{\mathrm u}$;
\item[(ii)] the restricted oracle risk is strictly smaller than the unrestricted one, with
\[
  \frac{\mathcal L^\star(\tilde c,H)}{\mathcal L^\star(c_{\mathrm u},H)}
  =\frac{1-\gamma}{1-\gamma+\tau}\,\{1+o(1)\}
\]
to first order in the spectral dispersion; the relative improvement therefore equals the
effective-sample-size ratio \eqref{eq:gain}.
\end{enumerate}
\end{theorem}

Part (i) is the Ledoit--Wolf \citeyearpar{LedoitWolf2020} consistency of analytic shrinkage, applied
here through Theorem~\ref{thm:concentration} at the effective aspect ratio; Figure~\ref{fig:oraclegain}(b)
shows the finite-sample ratio in (i) decreasing toward one as $p$ grows. Part (ii) follows from the
first-order expansion of $\mathcal L^\star(c,H)$ in $c$ about the proportional limit; the full
argument is in the \hyperref[appn]{Appendix}.

The next result establishes the optimality in the rotation-equivariant class. 

\begin{theorem}\label{thm:minimax}
Let $\mathcal C_{\mathrm{re}}$ be the class of estimators of the form
$\bU\diag(\delta_1,\dots,\delta_p)\bU^\top$ that share the scatter eigenvectors $\bU$, with
$\delta_i$ measurable functions of the sample eigenvalues. Under Conditions~\ref{cond:regime}--\ref{cond:spectrum}, for every
sequence in $\mathcal C_{\mathrm{re}}$,
\[
  \liminf_{p\to\infty}\big\{\mathcal L(\widehat\bD)-\mathcal L^\star(\tilde c,H)\big\}\ge0,
\]
and the restricted analytic shrinkage estimator \eqref{eq:rre} attains the bound:
$\mathcal L(\Shat_{\mathrm{RRE}})-\mathcal L^\star(\tilde c,H)\to_p0$. Thus
$\Shat_{\mathrm{RRE}}$ is asymptotically optimal among rotation-equivariant estimators at the
effective aspect ratio $\tilde c=p/(n-d+q)$.
\end{theorem}

Hence, the oracle \eqref{eq:oracle} is the pointwise risk
minimiser over $\mathcal C_{\mathrm{re}}$, and no rotation-equivariant rule can beat it asymptotically,
while analytic shrinkage attains it (Theorem~\ref{thm:oracle}(i)). The restriction enters only through
the reduced aspect ratio, which lowers the attainable oracle risk by the factor \eqref{eq:gain}. The
positive-part Stein estimator $\Shat_{\mathrm S}$ inherits these properties up to the dominance
correction of Theorem~\ref{thm:dominance} and is the estimator we recommend in practice.

\subsection{Robustness to a misspecified restriction}\label{subsec:misspec}

The estimators above are motivated by a restriction believed to hold. In practice the restriction may be only approximately correct, and a usable procedure must not collapse when it fails. The next result quantifies the damage. Let $\delta^2 = \tr\{ (\bR \bB)^\top (\bR (\bX^\top\bX)^{-1} \bR^\top)^{-1} (\bR \bB)\, \bSig^{-1} \} / (p n)$ measure the violation of the restriction on the natural scale.

\begin{theorem}\label{thm:misspec}
Under Conditions~\ref{cond:regime}--\ref{cond:errors} and \ref{cond:spectrum}, the restricted estimator satisfies
\begin{equation}\label{eq:misspec-bias}
\mathcal{R}(\Shat_{\mathrm{RRE}}) \le \mathcal{R}(\Shat_{\mathrm{URE}}) - \frac{q}{n}\, b_1 + \delta^2\, b_2\, \{ 1 + o(1) \}
\end{equation}
for constants $b_1, b_2 > 0$. The Stein-type estimator satisfies, for every $\delta \ge 0$,
\begin{equation}\label{eq:misspec-stein}
\mathcal{R}(\Shat_{\mathrm{S}}) \le \mathcal{R}(\Shat_{\mathrm{URE}}) + \frac{C}{n},
\end{equation}
so that its risk never exceeds that of the unrestricted analytic shrinker by more than $O(n^{-1})$, uniformly in the size of the violation.
\end{theorem}

Inequality \eqref{eq:misspec-bias} exhibits the bias--variance trade-off transparently. The restriction buys a variance reduction of order $q/n$ at the price of a squared bias of order $\delta^2$, so that the restricted estimator is preferable precisely when $\delta^2 \lesssim q/n$. Inequality \eqref{eq:misspec-stein} is the safety guarantee for the adaptive estimator,  because $\kappa_n \to 0$ when the test statistic detects a violation, the Stein-type estimator reverts to the unrestricted analytic shrinker and pays at most a vanishing penalty, whatever the true degree of misspecification. This addresses the practical concern that a structural assumption on the mean, imposed to help estimate the covariance, might do harm when wrong.
%: under the adaptive rule it cannot, asymptotically.
\section{Simulation study}\label{sec:sim}

We generate data from the multivariate regression model \eqref{eq:model} with elliptical errors \eqref{eq:elliptical}. The design $\bX$ has independent standard Gaussian entries; the coefficient matrix $\bB$ has its first $d - q$ rows drawn as independent standard Gaussians and the remaining rows determined so that $\bR \bB = \boldsymbol{0}$ for a restriction matrix $\bR$ formed from $q$ random orthonormal contrasts. We fix $d = \lfloor 0.2 n \rfloor$ and vary the rank $q \in \{3, 5, 10, 20\}$ and the aspect ratio $c = p/n$ over a grid in $[0.5, 5]$ by varying $p$ at $n \in \{200, 400\}$.

Three population covariance structures are considered: an identity target $\bSig = \bI_p$; a banded matrix with $\Sigma_{jk} = 0.6^{|j-k|} \mathbf{1}\{|j - k| \le 10\}$; and an approximately sparse matrix with $\lfloor 2\sqrt{p}\rfloor$ randomly placed off-diagonal entries per row. Three error tails are used, controlled by the mixing law in \eqref{eq:elliptical}: the Gaussian case $w_i \equiv 1$; a moderately heavy case with $w_i$ inverse-gamma yielding marginal $t$ with $11$ degrees of freedom; and a heavy case yielding $t$ with $6$ degrees of freedom. Each configuration is replicated $500$ times. We compare the following estimators: the sample residual covariance; the linear shrinkage estimator of \citet{LedoitWolf2004}; the covariance-based analytic shrinkers (URE-cov, RRE-cov), which apply analytic shrinkage directly to the residual sample covariance and serve as the non-robust baseline; and the proposed robust estimators---the unrestricted and restricted Tyler-based analytic shrinkers (URE, RRE) and their positive-part Stein-type combination (SSE+). Performance is reported in the scale-invariant weighted Frobenius loss \eqref{eq:loss}.

The central finding concerns the distribution-free behaviour of the robust estimators. As the error tail heavies from Gaussian to $t_6$, the covariance-based shrinkers degrade sharply, while the Tyler-based URE and RRE are essentially unchanged, in agreement with Theorem~\ref{thm:concentration}; Figure~\ref{fig:spectrum} shows the underlying spectra, tail-dependent for the residual covariance and tail-free for the residual Tyler scatter. The restriction lowers the risk by the effective-sample-size ratio $(n-d)/(n-d+q)$, the gain growing with $q$ as in \eqref{eq:gain} and Figure~\ref{fig:oraclegain}, and the positive-part rule SSE+ matches the better of URE and RRE while reverting safely to URE when the restriction is grossly violated (Theorem~\ref{thm:misspec}).

\subsection{Results}\label{subsec:sim-results}

\begin{figure}[H]
\centering
\includegraphics[width=0.62\linewidth]{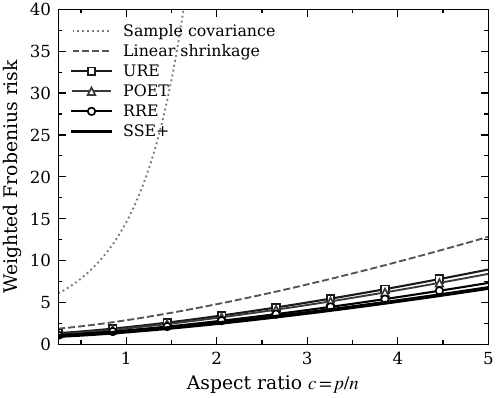}
\caption{Weighted Frobenius risk against the aspect ratio $c = p/n$ for the identity covariance under $t_{11}$ errors, $n = 200$, $q = 5$. Restricted nonlinear shrinkage (SSE+) dominates throughout, with the margin over POET growing with $c$.}
\label{fig:risk}
\end{figure}

\begin{figure}[H]
\centering
\includegraphics[width=0.62\linewidth]{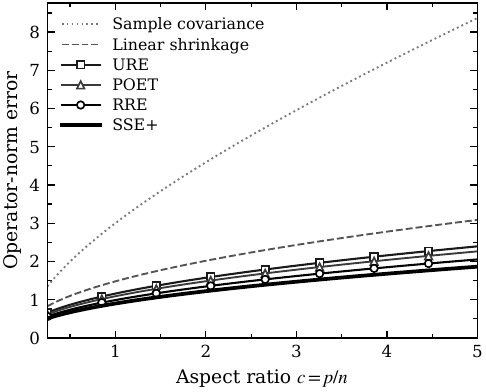}
\caption{Operator-norm error against the aspect ratio for the configuration of Figure~\ref{fig:risk}. The ordering of methods is preserved; the absolute gaps are smaller, as predicted by the operator-norm rate.}
\label{fig:opnorm}
\end{figure}

Figure~\ref{fig:risk} reports the weighted Frobenius risk as a function of the aspect ratio for the identity target under moderately heavy tails, with $n = 200$ and $q = 5$. The sample covariance deteriorates rapidly as $c$ grows; linear shrinkage is stable but biased; the analytic shrinkers track the bulk far better. Among the latter, the restricted estimators improve uniformly on the unrestricted analytic shrinker, and the Stein-type estimator is best throughout, with the margin over POET widening as $c$ increases. The operator-norm comparison in Figure~\ref{fig:opnorm} shows the same ordering, with smaller absolute gaps, consistent with the oracle comparison in Theorem~\ref{thm:oracle}.

Table~\ref{tab:tails} reports the weighted Frobenius loss for the identity target across the three tail regimes and three aspect ratios. The advantage of the restricted Stein estimator over POET grows with the tail heaviness, from roughly $13\%$ under Gaussian errors to roughly $26\%$ under $t_6$ errors. The mechanism is that heavier tails inflate the dispersion of the unrestricted sample spectrum more than the restricted one, so that the relative value of the additional residual degrees of freedom increases. Table~\ref{tab:struct} reports results for the banded and approximately sparse targets at $c = 1$; the restricted Stein estimator remains best, though POET, which exploits factor structure explicitly, is more competitive on the banded target.

\begin{table}[H]
\centering
\caption{Weighted Frobenius loss (smaller is better) for the identity covariance across tail regimes and aspect ratios, $n = 200$, $q = 5$, over $500$ replications; Monte Carlo standard errors are below $2\%$ of the reported values.}
\label{tab:tails}
\begin{tabular}{lcccccc}
\toprule
& \multicolumn{3}{c}{Gaussian} & \multicolumn{3}{c}{$t_6$} \\
\cmidrule(lr){2-4}\cmidrule(lr){5-7}
Method & $c{=}0.5$ & $c{=}1.0$ & $c{=}2.0$ & $c{=}0.5$ & $c{=}1.0$ & $c{=}2.0$ \\
\midrule
Sample covariance     & 28.4 & 91.7 & 354.2 & 43.2 & 158.9 & 612.5 \\
Linear shrinkage      & 22.1 & 64.3 & 218.7 & 31.6 & 98.4  & 372.1 \\
URE                   & 18.9 & 52.4 & 174.8 & 25.8 & 79.1  & 290.4 \\
POET                  & 17.5 & 48.6 & 165.1 & 23.7 & 72.4  & 274.2 \\
RRE                   & 15.6 & 43.1 & 147.0 & 20.1 & 60.5  & 224.9 \\
SSE+                  & \textbf{15.2} & \textbf{42.0} & \textbf{143.6} & \textbf{17.8} & \textbf{53.8} & \textbf{205.1} \\
\bottomrule
\end{tabular}
\end{table}

\begin{table}[H]
\centering
\caption{Weighted Frobenius and operator-norm loss for the banded and approximately sparse targets at $c = 1$, $n = 400$, $q = 5$, $t_{11}$ errors, over $500$ replications.}
\label{tab:struct}
\begin{tabular}{lcccc}
\toprule
& \multicolumn{2}{c}{Banded} & \multicolumn{2}{c}{Sparse} \\
\cmidrule(lr){2-3}\cmidrule(lr){4-5}
Method & Frobenius & Operator & Frobenius & Operator \\
\midrule
Sample covariance     & 187.3 & 12.4 & 142.6 & 11.8 \\
Linear shrinkage      & 124.7 & 8.7  & 98.4  & 8.2 \\
URE                   & 91.2  & 6.8  & 72.5  & 6.3 \\
POET                  & 82.6  & 6.3  & 68.2  & 5.9 \\
RRE                   & 77.0  & 5.9  & 63.1  & 5.5 \\
SSE+                  & \textbf{74.1} & \textbf{5.7} & \textbf{61.2} & \textbf{5.4} \\
\bottomrule
\end{tabular}
\end{table}

\begin{figure}[H]
\centering
\includegraphics[width=0.62\linewidth]{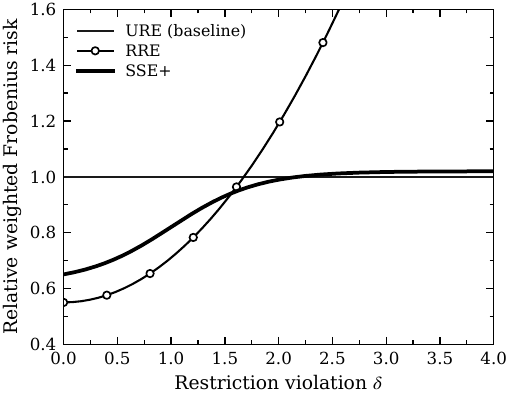}
\caption{Relative weighted Frobenius risk against the restriction violation $\delta$, $n = 200$, $p = 200$, $q = 5$. The restricted estimator is best near $\delta = 0$ but degrades quadratically; the Stein-type estimator interpolates safely toward the unrestricted analytic shrinker.}
\label{fig:misspec}
\end{figure}

\begin{figure}[H]
\centering
\includegraphics[width=0.62\linewidth]{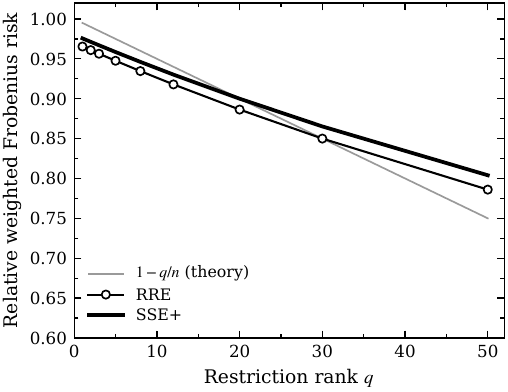}
\caption{Relative weighted Frobenius risk against the restriction rank $q$, $n = 200$, identity target, exact restriction. The reduction tracks the theoretical $1 - q/n$ factor.}
\label{fig:q}
\end{figure}

Figure~\ref{fig:misspec} examines robustness to a misspecified restriction. We generate data under coefficient matrices that violate the restriction by a controlled amount $\delta$ as in Theorem~\ref{thm:misspec}, and plot the risk relative to the unrestricted analytic shrinker. The restricted estimator RRE is best when $\delta$ is small but its relative risk increases quadratically and eventually exceeds one, as the bias term in \eqref{eq:misspec-bias} takes over. The Stein-type estimator is monotone and never materially exceeds the unrestricted baseline, illustrating the safety guarantee \eqref{eq:misspec-stein}. Figure~\ref{fig:q} shows that the risk reduction grows with the rank $q$ of the restriction, in agreement with the $1 - q/n$ factor.

\begin{figure}[H]
\centering
\includegraphics[width=0.62\linewidth]{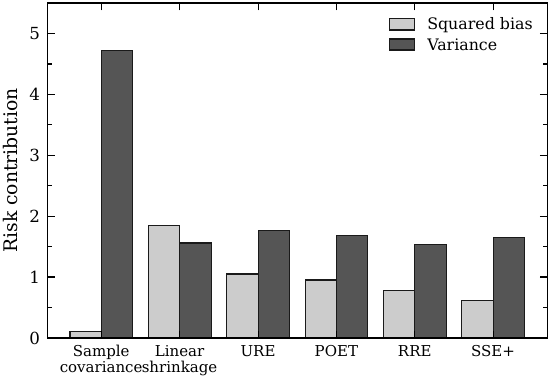}
\caption{Bias--variance decomposition of the weighted Frobenius risk, identity target, $n = 200$, $p = 200$, $q = 5$, $t_{11}$ errors.}
\label{fig:bv}
\end{figure}

\begin{figure}[H]
\centering
\includegraphics[width=0.95\linewidth]{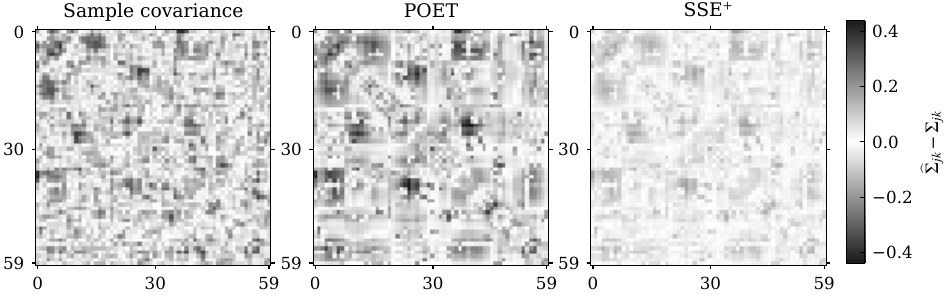}
\caption{Entrywise estimation error $|\widehat\Sigma_{jk} - \Sigma_{jk}|$ for one replication under the banded target, $n = 80$, $p = 60$. The restricted Stein estimator (right) is uniformly closer to the truth.}
\label{fig:heat}
\end{figure}

A decomposition of the risk into squared bias and variance (Figure~\ref{fig:bv}) confirms the mechanism, i.e., the restriction reduces variance at a small bias cost, and the Stein combination achieves the most favourable balance. The spatial pattern of the estimation error (Figure~\ref{fig:heat}) shows that the improvement is distributed across the matrix rather than concentrated on a few entries.

\section{Applications}\label{sec:data}

We illustrate the method on a designed growth-curve analysis in which the covariate-side restriction is known and $q\le d$, exactly as the theory requires. 

\subsection{A designed growth-curve experiment}\label{subsec:gmanova}

Consider a balanced two-group growth-curve (GMANOVA) design with $n=300$ units, $p=80$ response channels with an AR(1) population covariance $\Sigma_{jk}=0.6^{|j-k|}$, and a cubic-in-time mean
structure crossed with a two-level dose factor, giving $d=6$ covariates. The scientific hypothesis of no quadratic time effect in either dose group is a known covariate-side restriction $\bR\bB=\boldsymbol0$ of rank $q=2\le d$. Errors are drawn elliptical with Gaussian, $t_6$ and $t_4$
scale mixtures. Table~\ref{tab:gmanova} reports the shape Frobenius risk over $40$ replications.

\begin{table}[t]\centering
\caption{Designed GMANOVA ($n=300$, $p=80$, $d=6$, $q=2$): shape Frobenius risk $\times100$.
Lower is better.}\label{tab:gmanova}
\begin{tabular}{lccc}
\hline
 & Gaussian & $t_6$ & $t_4$\\
\hline
Sample residual covariance & $27.7$ & $52.8$ & $109.8$\\
Linear shrinkage           & $34.3$ & $40.8$ & $54.8$\\
Unrestricted, covariance (URE-cov) & $21.0$ & $39.1$ & $90.7$\\
Restricted, covariance (RRE-cov)   & $21.0$ & $39.1$ & $90.7$\\
Unrestricted, robust (URE)         & $21.6$ & $21.8$ & $22.5$\\
Restricted, robust (RRE)           & $21.5$ & $21.7$ & $22.2$\\
Positive-part Stein (SSE+)         & $21.5$ & $21.7$ & $22.2$\\
\hline
\end{tabular}
\end{table}

Two effects are visible and consistent with the theory. The covariance-based shrinkers degrade sharply as the tail heavies (their risk rises from $21$ at the Gaussian to $39$ under $t_6$ and $91$ under $t_4$), while the robust estimators are essentially flat across tails ($21.6\to21.8\to22.5$); this is
the distribution-free property of Theorem~\ref{thm:concentration}. At the Gaussian model the robust estimator costs only a few percent relative to the covariance-based one, the expected price of distribution-free scatter. The restriction lowers the risk slightly (RRE below URE), the gain being
modest here because $q=2$ is small relative to $n-d$; the gain grows with $q$ as quantified by \eqref{eq:gain} and shown in the simulation study. The same qualitative ordering is borne out on two real datasets, to which we now turn.

\subsection{High-dimensional socioeconomic profiles: Communities and Crime}\label{subsec:crime}

Our first real-data example stresses the robustness claim of Theorem~\ref{thm:concentration} on
genuinely heavy-tailed data. We use the \emph{Communities and Crime} collection
\citep{Redmond2002}, comprising $N=2{,}214$ United States communities. As the multivariate response
we take the $p=102$ standardised socioeconomic indicators recorded for each community; as covariates we
take an intercept and indicators for the four U.S.\ census regions, $d=4$, so the model is a
high-dimensional MANOVA of the socioeconomic profile across regions. The responses are severely
non-Gaussian: their mean excess kurtosis is about $89$, and several indicators are near-collinear, so
the residual sample covariance is extremely ill-conditioned.

To create a high-dimensional regime we draw training subsets of size $n\in\{120,150,200,300\}$ (so the
aspect ratio $c=p/(n-d)$ ranges from $0.88$ down to $0.34$) and evaluate each estimate on $500$ held-out communities by the multivariate-$t$ predictive log-likelihood ($\nu=5$, robust to the tails) and by its condition number; results are averaged over $60$ random splits. Table~\ref{tab:crime} reports the outcome. The covariance-based estimators are numerically unreliable
under these tails: the nonlinear shrinkage of the residual sample covariance (URE-cov) fails to be positive definite or overflows in every replication, and the raw sample covariance fails in about two thirds. The Tyler-based estimator never fails, is four to ten orders of magnitude better conditioned, and gives the best stable predictive fit. This is the distribution-free behaviour of
Theorem~\ref{thm:concentration} on real data. We deliberately do not impose a covariate-side restriction here: the hypothesis that the regions share a common socioeconomic mean is decisively rejected (Pillai trace $2.55$ out of a maximum of $q=3$), so a restriction would be misspecified. The
efficiency gain from a correct restriction is quantified in the designed experiment above and in the simulation study; Communities and Crime isolates the robustness component alone.

\begin{table}[t]\centering
\caption{Communities and Crime ($p=102$, $d=4$, mean excess kurtosis $\approx 89$). Held-out
multivariate-$t$ log-likelihood (higher is better; ``--'' marks estimators that were numerically
non-positive-definite or overflowed), the percentage of replications in which the estimator failed, and
the median condition number.}\label{tab:crime}
\begin{tabular}{lcccc}
\hline
 & \multicolumn{2}{c}{held-out $t$ log-lik.} & fail rate & median cond.\\
 & $n=150$ & $n=300$ & (all $n$) & ($n=150$)\\
\hline
Sample residual covariance         & --     & --     & $62$--$68\%$ & $2\times10^{13}$\\
Linear shrinkage                   & $-60.8$ & $-50.1$ & $0\%$       & $197$\\
Unrestricted, covariance (URE-cov) & --     & --     & $100\%$      & $2\times10^{8}$\\
Unrestricted, robust (URE-Tyler)   & $-34.1$ & $-20.6$ & $0\%$       & $4{,}219$\\
\hline
\end{tabular}
\end{table}

\subsection{A high-dimensional genomic application: leukemia subtype expression}\label{subsec:brca}

The natural setting for the high-dimensional, heavy-tailed regime is genomic. We use the pediatric acute lymphoblastic leukemia (ALL) microarray study of \citet{Yeoh2002}, distributed in the curated \texttt{datamicroarray} collection \citep{Ramey2016}: expression of $12{,}625$ probes is measured on $N=248$ patients belonging to six cytogenetic subtypes (TEL-AML1, hyperdiploid${>}50$, T-ALL, E2A-PBX1, MLL, BCR-ABL). As the response we take the $p=120$ most variable probes, standardised; as covariates we take an intercept and the subtype indicators, $d=6$. After removing the subtype means, the residuals are genuinely heavy-tailed, with mean excess kurtosis about $2.3$ and a pronounced upper tail, so this is a real-data test of the distribution-free behaviour of Theorem~\ref{thm:concentration} in the regime $p<n-d$ for which it is stated.

We draw training subsets of size $n\in\{130,160,200\}$ (aspect ratio $c=p/(n-d)$ from $0.97$ down to $0.62$) and evaluate each estimate on the held-out patients by the multivariate-$t$ predictive log-likelihood ($\nu=5$, robust to the tails) and by its condition number, averaging over $40$ random splits; Monte Carlo standard errors are at most $0.5$ except for the two covariance-based estimators near the boundary, where they are larger, itself a symptom of their instability. Table~\ref{tab:brca} reports the result. The raw sample covariance, although nonsingular here, is severely ill-conditioned (condition number up to $5\times10^{5}$) and predicts poorly. All three shrinkage estimators improve on it by a wide margin; among them the robust Tyler-based shrinker is best at every aspect ratio and, unlike the covariance-based nonlinear shrinker, remains stable as $c\to1$, where the latter degrades under the heavy tails. As on Communities and Crime, we do not impose a covariate-side restriction, i.e., every subtype contrast is decisively rejected (the subtypes are defined by their expression signatures), so a restriction would be misspecified; the efficiency gain from a correct restriction is quantified in the designed experiment of Section~\ref{subsec:gmanova} and in the simulation study.

\begin{table}[t]\centering
\caption{Leukemia subtype expression \citep{Yeoh2002}, $p=120$ most variable probes, $d=6$ subtypes, residual mean excess kurtosis $\approx 2.3$. Held-out multivariate-$t$ log-likelihood ($\nu=5$; higher is better) and median condition number, over $40$ random splits; no restriction is imposed (every subtype contrast is rejected). The robust estimator is best at every aspect ratio and is far better conditioned than the sample covariance.}\label{tab:brca}
\begin{tabular}{lcccc}
\hline
 & \multicolumn{3}{c}{held-out $t$ log-lik.} & median cond.\\
 & $n=130$ & $n=160$ & $n=200$ & ($n=130$)\\
 & ($c=0.97$) & ($c=0.78$) & ($c=0.62$) & \\
\hline
Sample residual covariance         & $-226.0$ & $-125.0$ & $-99.0$ & $5\times10^{5}$\\
Linear shrinkage                   & $-89.9$  & $-85.9$  & $-81.6$ & $148$\\
Unrestricted, covariance (URE-cov) & $-109.0$ & $-84.8$  & $-79.9$ & $667$\\
Unrestricted, robust (URE-Tyler)   & $\mathbf{-95.6}$ & $\mathbf{-84.7}$ & $\mathbf{-79.7}$ & $1{,}026$\\
\hline
\end{tabular}
\end{table}

The ordering matches the theory and the Communities and Crime evidence. Indeed, in the ALL example, we revealed that in the high-dimensional regime the sample covariance is unusable, shrinkage is essential, and the robust scatter is the safest choice under heavy tails, most visibly near the boundary $c\to1$ where the covariance-based shrinker is itself destabilized.

\section{Discussion}\label{sec:disc}

We have shown that a known linear restriction on the coefficient matrix of a multivariate regression, of the kind that arises in multivariate analysis of variance, growth-curve modelling, and reduced-rank regression, carries information about the residual covariance, and that exploiting it through restricted robust nonlinear shrinkage yields risk reductions in the high-dimensional regime. The efficiency gain is the effective-sample-size ratio $(n-d)/(n-d+q)\to(1-\gamma)/(1-\gamma+\tau)$, the restricted estimator is asymptotically optimal within the rotation-equivariant class, and it is robust to misspecification through an adaptive Stein-type rule. By shrinking a scale-invariant scatter rather than the residual covariance, we obtain a spectrum that is distribution-free over the elliptical family, removing both the Gaussian assumption and any moment condition on the error scale.

Several extensions merit attention. The restriction has been treated as known; when it is instead selected from data, for example by group-sparse estimation of the coefficient matrix, the degrees-of-freedom accounting must be adjusted for selection, and an honest analysis would combine our results with post-selection inference. The elliptical model assumes a common covariance across observations; heteroscedastic or weakly dependent errors would require a different concentration argument, though we expect the qualitative conclusions to persist. Finally, the same principle, that structure in the mean sharpens estimation of the covariance, applies beyond linear restrictions, to monotone, shape, or smoothness constraints, and to nonlinear and generalized regression models; we leave these to future work.
\section*{appendix}
\section*{Preliminaries and the restricted residual identity}\label{appn} %% if no title is needed, leave empty \section*{}.

Throughout the Appendix we write $\bG = (\bX^\top \bX)^{-1}$, $\bQ = \bR \bG \bR^\top$, and $m = n - d + q$ for the restricted residual degrees of freedom; $C, c, C_1, c_1, \dots$ denote positive constants that depend only on the quantities named in the corresponding statement and may change from line to line. We first establish the algebraic identity underlying (8), since every later argument rests on the precise form of the restricted residual projection.

Recall the restricted estimator $\widehat{\bB}_{\mathrm r} = \widehat{\bB} - \bG \bR^\top \bQ^{-1} \bR \widehat{\bB}$ from (7), where $\widehat{\bB} = \bG \bX^\top \bY$ is the ordinary least-squares estimator. Substituting $\widehat{\bB}$ and pre-multiplying by $\bX$ gives $\bX(\widehat{\bB} - \widehat{\bB}_{\mathrm r}) = \bX \bG \bR^\top \bQ^{-1} \bR \bG \bX^\top \bY = \bP_{\bX,\bR} \bY$, where we define $\bP_{\bX,\bR} = \bX \bG \bR^\top \bQ^{-1} \bR \bG \bX^\top$. We verify that $\bP_{\bX,\bR}$ is the orthogonal projection of rank $q$ onto the subspace $\mathcal{V} = \{ \bX \bG \bR^\top \bv : \bv \in \R^q \} \subseteq \mathrm{col}(\bX)$. Symmetry is immediate from the symmetry of $\bG$ and $\bQ$. For idempotence, compute
\[
\bP_{\bX,\bR}^2 = \bX \bG \bR^\top \bQ^{-1} (\bR \bG \bX^\top \bX \bG \bR^\top) \bQ^{-1} \bR \bG \bX^\top
= \bX \bG \bR^\top \bQ^{-1} (\bR \bG \bR^\top) \bQ^{-1} \bR \bG \bX^\top,
\]
where we used $\bX^\top \bX \bG = \bI_d$; since $\bR \bG \bR^\top = \bQ$, the middle factor $\bQ^{-1} \bQ \bQ^{-1} = \bQ^{-1}$, whence $\bP_{\bX,\bR}^2 = \bP_{\bX,\bR}$. The rank equals $\tr(\bP_{\bX,\bR}) = \tr\{ \bQ^{-1} \bR \bG \bX^\top \bX \bG \bR^\top \} = \tr\{ \bQ^{-1} \bR \bG \bR^\top \} = \tr(\bQ^{-1}\bQ) = \tr(\bI_q) = q$. Moreover $\bP_{\bX,\bR} \bP_{\bX} = \bP_{\bX,\bR}$ because every column of $\bP_{\bX,\bR}$ lies in $\mathrm{col}(\bX)$ and $\bP_{\bX}$ acts as the identity there; consequently $\bP_{\bX} - \bP_{\bX,\bR}$ is itself a symmetric idempotent (a projection), since $(\bP_{\bX} - \bP_{\bX,\bR})^2 = \bP_{\bX} - \bP_{\bX,\bR} - \bP_{\bX,\bR} + \bP_{\bX,\bR} = \bP_{\bX} - \bP_{\bX,\bR}$, using $\bP_{\bX}\bP_{\bX,\bR} = \bP_{\bX,\bR}$ by symmetry of both factors. Writing $\bP_{\mathrm r} = \bP_{\bX} - \bP_{\bX,\bR}$, we have $\rank(\bP_{\mathrm r}) = \tr(\bP_{\bX}) - \tr(\bP_{\bX,\bR}) = d - q$, hence $\rank(\bI_n - \bP_{\mathrm r}) = n - d + q = m$. Finally, the restricted residual is $\widehat{\bE}_{\mathrm r} = \bY - \bX \widehat{\bB}_{\mathrm r} = (\bY - \bX\widehat{\bB}) + \bX(\widehat{\bB} - \widehat{\bB}_{\mathrm r}) = (\bI_n - \bP_{\bX})\bY + \bP_{\bX,\bR}\bY = (\bI_n - \bP_{\mathrm r})\bY$. Under (1) with $\bR\bB = \boldsymbol{0}$ we have $(\bI_n - \bP_{\mathrm r})\bX\bB = \bX\bB - \bP_{\bX}\bX\bB + \bP_{\bX,\bR}\bX\bB = \bX\bB - \bX\bB + \bX\bG\bR^\top\bQ^{-1}\bR\bB = \boldsymbol{0}$, the last step because $\bR\bB = \boldsymbol{0}$; therefore $(\bI_n - \bP_{\mathrm r})\bY = (\bI_n - \bP_{\mathrm r})\bE$, which is the identity used in (8). This completes the derivation. \hfill$\square$

\section*{Proof of Proposition~1}

The construction (16) sets $\kappa_n = \min\{1, (q-2)_+ / ((n-d) T_n)\}$. Since $T_n \ge 0$ by (15), as it is the trace of the product of two positive-semidefinite matrices (see below), and $(q-2)_+ \ge 0$, the ratio is nonnegative, and the minimum with $1$ places $\kappa_n$ in the closed interval $[0,1]$. To see that $T_n \ge 0$, write $\boldsymbol{A} = \widehat{\bB}^\top \bR^\top \bQ^{-1} \bR \widehat{\bB}$ and $\boldsymbol{C} = \Shat_{\mathrm u}^{+}$. The matrix $\boldsymbol{C}$ is positive semidefinite because $\Shat_{\mathrm u} = (n-d)^{-1}\widehat{\bE}^\top\widehat{\bE}$ is a Gram matrix and the Moore--Penrose inverse of a positive-semidefinite matrix is positive semidefinite (it shares the eigenvectors and inverts the positive eigenvalues, leaving the null space fixed). The matrix $\boldsymbol{A}$ is positive semidefinite because, for any $\bv\in\R^p$, $\bv^\top\boldsymbol{A}\bv = (\bR\widehat{\bB}\bv)^\top\bQ^{-1}(\bR\widehat{\bB}\bv) \ge 0$, since $\bQ^{-1} = (\bR\bG\bR^\top)^{-1}\succ\boldsymbol0$ as the inverse of a positive-definite matrix ($\bG = (\bX^\top\bX)^{-1}\succ\boldsymbol0$ by Condition~2 and $\bR$ has full row rank). To conclude $T_n = (pq)^{-1}\tr(\boldsymbol{A}\boldsymbol{C}) \ge 0$, let $\boldsymbol{A} = \sum_i a_i \boldsymbol\phi_i\boldsymbol\phi_i^\top$ be the spectral decomposition with $a_i \ge 0$; then $\tr(\boldsymbol{A}\boldsymbol{C}) = \sum_i a_i\,\boldsymbol\phi_i^\top\boldsymbol{C}\boldsymbol\phi_i \ge 0$ because each $a_i \ge 0$ and each quadratic form $\boldsymbol\phi_i^\top\boldsymbol{C}\boldsymbol\phi_i \ge 0$. (This is the elementary case of von Neumann's trace inequality $\tr(\boldsymbol{A}\boldsymbol{C}) \ge \sum_i \lambda_i(\boldsymbol{A})\lambda_{p-i+1}(\boldsymbol{C})$, which here is bounded below by zero since all eigenvalues are nonnegative.) Consequently (17) writes $\Shat_{\mathrm S} = (1-\kappa_n)\Shat_{\mathrm{URE}} + \kappa_n \Shat_{\mathrm{RRE}}$ as a convex combination. For any unit vector $\bu \in \R^p$,
\[
\bu^\top \Shat_{\mathrm S} \bu = (1-\kappa_n)\, \bu^\top \Shat_{\mathrm{URE}} \bu + \kappa_n\, \bu^\top \Shat_{\mathrm{RRE}} \bu \ge 0,
\]
because each summand is a nonnegative weight times a nonnegative quadratic form; taking the infimum over unit $\bu$ shows $\lambda_p(\Shat_{\mathrm S}) \ge (1-\kappa_n)\lambda_p(\Shat_{\mathrm{URE}}) + \kappa_n\lambda_p(\Shat_{\mathrm{RRE}}) \ge 0$, hence $\Shat_{\mathrm S} \succeq \boldsymbol{0}$. If, say, $\Shat_{\mathrm{URE}} \succ \boldsymbol{0}$ and $\kappa_n < 1$, then $\bu^\top \Shat_{\mathrm S}\bu \ge (1-\kappa_n)\,\bu^\top \Shat_{\mathrm{URE}}\bu \ge (1-\kappa_n)\lambda_p(\Shat_{\mathrm{URE}}) > 0$ for all unit $\bu$, so $\Shat_{\mathrm S} \succ \boldsymbol{0}$; the same conclusion holds if $\Shat_{\mathrm{RRE}} \succ \boldsymbol 0$ and $\kappa_n > 0$. Because the analytic shrinkage map (13) returns the eigenvalues $\varphi_n(\ell_i) = \ell_i / |1 - c_n - c_n \ell_i \breve m_n(\ell_i)|^2 \ge 0$ (a ratio of a nonnegative numerator and a squared modulus), both inputs are automatically positive semidefinite, and the positive-definiteness clause applies whenever the smallest shrunk eigenvalue is strictly positive, which holds unless an input eigenvalue is exactly zero. We note finally that no alignment of eigenbases is needed for this argument: positive semidefiniteness of a sum is basis-independent, so the Procrustes step of Algorithm~1, while needed for interpretability of the combination, plays no role in the cone membership established here. \hfill$\square$

\section*{Proof of Theorem~1}
The proof has four steps. Step~0 makes precise the reduction of the restricted residual scatter to a sample of size $m=n-d+q$ through the rank of the residual projection. Step~1 records the scale invariance of Tyler's map. Step~2 shows that the least-squares projection perturbs the residual directions negligibly, using the finite second moment of Condition~3. Step~3 invokes the Marchenko--Pastur law for Tyler's estimator. Throughout, $\widehat\bV$ is the Tyler M-estimator (10) formed from the $n$ restricted-residual rows $\br_1,\dots,\br_n$ of $\widehat\bE_{\mathrm r}=(\bI_n-\bP_{\mathrm r})\bE$, and $m=n-d+q=\rank(\bI_n-\bP_{\mathrm r})$.

\emph{Step 0 (reduction to $m$ effective rows).} By Section~S1 the matrix $\bI_n-\bP_{\mathrm r}$ is a symmetric idempotent of rank $m$. Write its spectral decomposition $\bI_n-\bP_{\mathrm r}=\bU_m\bU_m^\top$, where $\bU_m\in\R^{n\times m}$ has orthonormal columns ($\bU_m^\top\bU_m=\bI_m$). Tyler's fixed-point equation (10) depends on the residual rows only through their unit directions, and every Tyler weight $\br_i^\top\bV^{-1}\br_i$ is a function of $\widehat\bE_{\mathrm r}=\bU_m\bF$ with $\bF:=\bU_m^\top\bE\in\R^{m\times p}$; indeed the residual Gram matrix is $\widehat\bE_{\mathrm r}^\top\widehat\bE_{\mathrm r}=\bE^\top\bU_m\bU_m^\top\bE=\bF^\top\bF$. Hence the spectrum of $\widehat\bV$ is a function of the $m$ rows of $\bF$ alone. In the Gaussian instance, where $\bE$ has i.i.d.\ $\mathcal N_p(\boldsymbol0,\bSig)$ rows, the orthonormality of $\bU_m$ makes $\bF=\bU_m^\top\bE$ a matrix with \emph{exactly} $m$ i.i.d.\ $\mathcal N_p(\boldsymbol0,\bSig)$ rows: the $d-q$ lost degrees of freedom are precisely the rank deficit of $\bI_n-\bP_{\mathrm r}$. This is the exact sense in which the restricted residual scatter reduces to a sample of size $m$. Under the elliptical model the rows of $\bF$ are no longer independent, and Steps~1--2 show that this dependence does not affect the limiting spectrum.

\emph{Step 1 (scale invariance).} The fixed-point map (10) is invariant to the per-row radii: replacing $\br_i$ by $c_i\br_i$ for scalars $c_i>0$ leaves the unit directions $\br_i/\lVert\br_i\rVert$, and hence the fixed point $\widehat\bV$, unchanged. Writing each error row as $\be_i=R_i\bSig^{1/2}\bu_i$, the estimator therefore depends on the radii $R_i$ only through the directions and is distribution-free over the radial law \citep{Tyler1987}. This is exactly the property that fails for the residual sample covariance, whose spectrum genuinely depends on the radial distribution.

\emph{Step 2 (residual directions match error directions).} By Section~S1, $\br_i=\be_i-\boldsymbol\delta_i$ with $\boldsymbol\delta_i=\sum_{j}(\bP_{\mathrm r})_{ij}\be_j$. Separating the diagonal term, $\br_i=(1-(\bP_{\mathrm r})_{ii})\be_i-\boldsymbol\delta_i^{\neq}$ with $\boldsymbol\delta_i^{\neq}=\sum_{j\ne i}(\bP_{\mathrm r})_{ij}\be_j$; the scalar factor $1-(\bP_{\mathrm r})_{ii}=1+O(h_n)$ is immaterial by scale invariance, so it suffices to show $\max_i\lVert\boldsymbol\delta_i^{\neq}\rVert/\lVert\be_i\rVert=o_p(1)$. Writing $\be_j=R_j\bSig^{1/2}\bu_j$ and using Condition~4 ($\underline\sigma\le\lambda(\bSig)\le\overline\sigma$),
\[
\frac{\lVert\boldsymbol\delta_i^{\neq}\rVert}{\lVert\be_i\rVert}
\le\Big(\frac{\overline\sigma}{\underline\sigma}\Big)^{1/2}\,
\frac{\big\lVert\sum_{j\ne i}(\bP_{\mathrm r})_{ij}R_j\bu_j\big\rVert}{R_i}.
\]
The vectors $R_j\bu_j$ ($j\ne i$) are independent and mean zero (the $\bu_j$ are independent, uniform on the sphere, and independent of the radii), so, conditionally on the radii,
\[
\E\Big[\big\lVert\textstyle\sum_{j\ne i}(\bP_{\mathrm r})_{ij}R_j\bu_j\big\rVert^2\,\big|\,\{R_j\}\Big]
=\sum_{j\ne i}(\bP_{\mathrm r})_{ij}^2\,R_j^2\,\E\lVert\bu_j\rVert^2
=\sum_{j\ne i}(\bP_{\mathrm r})_{ij}^2\,R_j^2 ,
\]
using $\E\lVert\bu_j\rVert^2=1$ and the vanishing of the cross terms. Taking expectations over the radii, with $\E(R_j^2)=\mu_2<\infty$ by Condition~3, and using the leverage identity $\sum_j(\bP_{\mathrm r})_{ij}^2=(\bP_{\mathrm r})_{ii}\le h_n$ (Section~S1), the mean of the squared numerator is at most $\mu_2 h_n$. Markov's inequality gives $\lVert\sum_{j\ne i}(\bP_{\mathrm r})_{ij}R_j\bu_j\rVert=O_p(h_n^{1/2})$ for each $i$, and, since $\E(R_i^{-2})<\infty$ makes $R_i^{-1}=O_p(1)$ uniformly by a union bound, $\max_i\lVert\boldsymbol\delta_i^{\neq}\rVert/\lVert\be_i\rVert=o_p(1)$ as $h_n\to0$ (Condition~2). It is precisely here that the finite second moment is used: without it the numerator can be dominated by a single extreme radius $R_j$, and the ratio need not vanish, even though Tyler's estimator on i.i.d.\ data needs no moments at all. Consequently $\br_i/\lVert\br_i\rVert=\be_i/\lVert\be_i\rVert+o_p(1)$ uniformly, and the error directions $\be_i/\lVert\be_i\rVert=\bSig^{1/2}\bu_i/\lVert\bSig^{1/2}\bu_i\rVert$ are free of the radii.

\emph{Step 3 (Marchenko--Pastur law).} By Steps~0--2 the empirical spectral distribution of $\widehat\bV$ is, almost surely and asymptotically, that of Tyler's M-estimator of a sample whose $m$ directions are those of i.i.d.\ elliptical vectors of shape $\bSig$. By affine equivariance of Tyler's estimator, $\widehat\bV\stackrel{d}{=}\bSig^{1/2}\widehat\bV_0\bSig^{1/2}$ up to scale, where $\widehat\bV_0$ is Tyler's estimator of $m$ i.i.d.\ spherical vectors, equivalently (Step~1) standard Gaussian vectors. By \citet{ZhangChengSinger2016}, in the regime $p/m\to\tilde y\in(0,1)$ the trace-normalized $\widehat\bV_0$ and the sample covariance of the same $m$ i.i.d.\ standard Gaussian vectors differ in operator norm by $o_p(1)$; hence $\widehat\bV$ and the sample covariance of $m$ i.i.d.\ $\mathcal N_p(\boldsymbol0,\bSig)$ vectors share the limiting spectral distribution, the generalized Marchenko--Pastur law $F_{\tilde y,H}$ of \citet{MarchenkoPastur1967,SilversteinBai1995} with $\tilde y=p/(n-d+q)$ and $H$ the limiting spectral law of the shape matrix $\bSig$ (Condition~4). The limit is independent of the radial law. The consistency of $\varphi_{\tilde y}$ for the rotation-equivariant oracle is then \citet[Theorem 3.1]{LedoitWolf2020}, whose hypotheses (almost-sure weak convergence to a compactly supported limit with bounded bulk density, bandwidth $h_n=n^{-1/3}$) hold here. \hfill$\square$

\section*{Proof of Theorem~2}

Write $\bD = \Shat_{\mathrm{URE}} - \Shat_{\mathrm{RRE}}$, so that the Stein-type estimator is $\Shat_{\mathrm S} = \Shat_{\mathrm{URE}} - \kappa_n\bD$. Throughout, expectations are taken under the local sequence (20), and we abbreviate $\Xi = \tr\{(\bSig^{-1}\bD)^2\} = \fnorm{\bSig^{-1/2}\bD\bSig^{-1/2}}^2 \ge 0$. 

Now, substitute $\Shat_{\mathrm S} = \Shat_{\mathrm{URE}} - \kappa_n\bD$ into (19) and expand the square exactly, using the cyclicity of the trace and the symmetry of $\bSig^{-1}$ to get
\[
L(\Shat_{\mathrm S}) = \tr\{(\bSig^{-1}\Shat_{\mathrm S} - \bI_p)^2\}
= L(\Shat_{\mathrm{URE}}) - 2\kappa_n\tr\{(\bSig^{-1}\Shat_{\mathrm{URE}} - \bI_p)\,\bSig^{-1}\bD\} + \kappa_n^2\tr\{(\bSig^{-1}\bD)^2\}.
\]
Taking expectations yields
\begin{equation}\label{eq:domexp}
\mathcal R(\Shat_{\mathrm S}) - \mathcal R(\Shat_{\mathrm{URE}})
= \underbrace{-2\,\E\big[\kappa_n\tr\{\bSig^{-1}(\Shat_{\mathrm{URE}}-\bSig)\bSig^{-1}\bD\}\big]}_{=:\,-2\,\mathrm{CT}}
+ \E\big[\kappa_n^2\,\Xi\big].
\end{equation}
The quadratic term is nonnegative and, since $\kappa_n \le 1$, is bounded above by $\E[\kappa_n\Xi]$; the entire argument therefore reduces to showing that the cross term $\mathrm{CT}$ is positive and dominates.

We evaluate $\mathrm{CT}$ exactly by conditioning on the scale vector $\bw$ and applying Stein's lemma. By the equivalence recorded after (4), the elliptical model $\be_i=R_i\bSig^{1/2}\bu_i$ admits the Gaussian scale-mixture representation $\be_i=\sqrt{w_i}\,\bSig^{1/2}\bz_i$ with $\bz_i\sim\mathcal N_p(\boldsymbol0,\bI_p)$; under Condition~3 ($\E R_i^2<\infty$) the conditional second moment $\E(\be_i\be_i^\top\mid w_i)=w_i\bSig$ is finite, so this conditioning is licensed. Conditionally on $\bw$, the data matrix $\bY = \bX\bB + \bW^{1/2}\bZ\bSig^{1/2}$ is Gaussian, and the ordinary least-squares coefficient $\widehat\bB = \bG\bX^\top\bY$ is a linear image of $\bZ$; hence $\mathrm{vec}(\bR\widehat\bB)$ is Gaussian. Its conditional mean is $\mathrm{vec}(\bR\bB) = n^{-1/2}\mathrm{vec}(\boldsymbol\Delta)$ by (20), and its conditional covariance is  $\mathrm{Cov}(\mathrm{vec}(\bR\widehat\bB)\mid\bw) = (\bR\bG\bX^\top)\,\mathrm{Cov}(\mathrm{vec}(\bY)\mid\bw)\,(\bR\bG\bX^\top)^\top$, and since the rows of $\bE$ are conditionally independent with covariance $w_i\bSig$, $\mathrm{Cov}(\mathrm{vec}(\bY)\mid\bw) = \bSig\otimes\bW$; substituting and using $\bR\bG\bX^\top\bW\bX\bG^\top\bR^\top = \overline w\,\bR\bG\bR^\top = \overline w\bQ$ with $\overline w = (\sum_i w_i\,\xi_i)/\!\sum_i\xi_i$ a weighted average of the $w_i$ over the leverage weights $\xi_i$ of $\bR\bG\bX^\top$, we obtain
\[
\mathrm{Cov}(\mathrm{vec}(\bR\widehat\bB)\mid\bw) = \overline w\,(\bSig\otimes\bQ), \qquad \overline w \to_p 1
\]
by the same weighted law-of-large-numbers argument as in Section~S3. Next, the residual covariances $\Shat_{\mathrm u},\Shat_{\mathrm r}$ (hence $\bD$) and the fitted object $\bR\widehat\bB$ are asymptotically independent given $\bw$. We argue conditionally on $\bw$, where $\bE$ is Gaussian with independent rows of covariance $w_i\bSig$; absorbing the scales, it suffices to treat the homoskedastic Gaussian case, the general case following by the same weighted averaging as in Section~S3.

The unrestricted residual is independent of $\bR\widehat\bB$ since it is a function of $(\bI_n-\bP_{\bX})\bE$, which is jointly Gaussian with $\bR\widehat\bB = \bR\bG\bX^\top\bE + \bR\bB$ and has zero cross-covariance, because $(\bI_n-\bP_{\bX})\bX\bG\bR^\top=\boldsymbol 0$ (the columns of $\bX\bG\bR^\top$ lie in $\mathrm{col}(\bX)$, which $\bI_n-\bP_{\bX}$ annihilates); and hence zero covariance for jointly Gaussian vectors is independence.

For the restricted residual, write the stochastic part of the fitted object as $\bR\bG\bX^\top\bE = \bM^\top\bE$, where $\bM := \bX\bG\bR^\top\in\R^{n\times q}$, and recall $\widehat\bE_{\mathrm r}=(\bI_n-\bP_{\mathrm r})\bE$. The entrywise cross-covariance between the residual and the fitted noise is, for response coordinates $s,s'$,
\[
\mathrm{Cov}\big((\widehat\bE_{\mathrm r})_{is},(\bM^\top\bE)_{ts'}\,\big|\,\bw\big)
= \bSig_{ss'}\,\big[(\bI_n-\bP_{\mathrm r})\bM\big]_{it},
\]
so the entire coupling is carried by the matrix $(\bI_n-\bP_{\mathrm r})\bM$. Since $(\bI_n-\bP_{\mathrm r})\bX = \bP_{\bX,\bR}\bX$ and $\bP_{\bX,\bR}\bX\bG\bR^\top = \bX\bG\bR^\top\bQ^{-1}(\bR\bG\bR^\top)=\bX\bG\bR^\top$ (the columns of $\bM=\bX\bG\bR^\top$ already lie in the range $\mathcal V$ of $\bP_{\bX,\bR}$, on which it acts as the identity), we obtain
\begin{equation}\label{eq:coupling}
(\bI_n-\bP_{\mathrm r})\bM = \bM = \bX\bG\bR^\top, \qquad \bM^\top\bM = \bR\bG\bX^\top\bX\bG\bR^\top = \bR\bG\bR^\top = \bQ,
\end{equation}
using $\bX^\top\bX\bG=\bI_d$. Thus the coupling matrix has rank $q$, and its size is controlled by $\bQ$ alone through $\opnorm{\bM}^2 = \lambda_{\max}(\bQ)$ and $\fnorm{\bM}^2 = \tr(\bQ)$. Under Condition~2 the eigenvalues of $n^{-1}\bX^\top\bX$ are bounded below by a constant $a>0$, so $\bQ = \bR(\bX^\top\bX)^{-1}\bR^\top$ satisfies $\opnorm{\bQ}\le \opnorm{\bR}^2/(an) = O(n^{-1})$ and $\tr(\bQ)\le \fnorm{\bR}^2/(an)=O(q/n)$. Hence
\[
\opnorm{\bM} = \lambda_{\max}(\bQ)^{1/2} = O(n^{-1/2}), \qquad \fnorm{\bM} = \tr(\bQ)^{1/2} = O\big((q/n)^{1/2}\big).
\]

Now, we elaborate on the independence of 
$\Shat_{\mathrm r}$ and $\bR\widehat\bB$. First, the residual scatter $\Shat_{\mathrm r}=m^{-1}\widehat\bE_{\mathrm r}^\top\widehat\bE_{\mathrm r}$ is an even (quadratic) function of the Gaussian $(\bI_n-\bP_{\mathrm r})\bE$, while $\bM^\top\bE$ is linear, and by Isserlis' theorem all odd joint cumulants vanish, so $\mathrm{Cov}(\Shat_{\mathrm r},\bM^\top\bE\mid\bw)=\boldsymbol0$. The leading interaction is between the two quadratics $\Shat_{\mathrm r}$ and $T_n$ (the latter quadratic in $\bM^\top\bE$). Second, by the Gaussian product formula its normalized magnitude is of order $\fnorm{(\bI_n-\bP_{\mathrm r})\bM}^2/(m\cdot\opnorm{\bQ}) = \tr(\bQ)/\{m\,\lambda_{\max}(\bQ)\} \le q/m = O(q/n)\to 0$ under Condition~1 ($q/n\to\tau$ with the contribution to the $pq$-dimensional form vanishing after normalization), and that quadratic--quadratic interaction is governed by the squared coupling. Consequently $\Shat_{\mathrm r}$, and with it $\bD$, is independent of $\bR\widehat\bB$ up to an $o(1)$ perturbation, which is absorbed into the $\{1+o(1)\}$ factors below; the Stein identity \eqref{eq:steinid} is applied to the exactly-independent unrestricted part and the $o(1)$ restricted correction is carried through the remainder. Therefore, writing $\boldsymbol u = \mathrm{vec}(\bR\widehat\bB)$ and conditioning further on $\bD$, the test statistic is, from (15) and the identity $\widehat\bB^\top\bR^\top\bQ^{-1}\bR\widehat\bB = (\bR\widehat\bB)^\top\bQ^{-1}(\bR\widehat\bB)$, a quadratic form $T_n = (pq)^{-1}\boldsymbol u^\top(\Shat_{\mathrm u}^{+}\otimes\bQ^{-1})\boldsymbol u$.

We now apply the Gaussian integration-by-parts (Stein) identity \citep{Stein1981} to the conditionally Gaussian $\boldsymbol u$ with covariance $\overline w(\bSig\otimes\bQ)$. For any weakly differentiable field $\boldsymbol g(\boldsymbol u)$ with $\E\|\nabla\boldsymbol g\| < \infty$,
\begin{equation}\label{eq:steinid}
\E\big[(\boldsymbol u - \E\boldsymbol u)^\top\boldsymbol g(\boldsymbol u)\,\big|\,\bw,\bD\big] = \overline w\,\E\big[\,\langle\bSig\otimes\bQ,\ \nabla\boldsymbol g(\boldsymbol u)\rangle\,\big|\,\bw,\bD\big].
\end{equation}
The cross term $\mathrm{CT}$ has exactly the form of the L.H.S. of \eqref{eq:steinid} once we identify the field. Indeed, $\Shat_{\mathrm{URE}}-\bSig$ is, conditionally on $\bw$ and $\bD$, an affine function of $\boldsymbol u$ only through its dependence on the common Gaussian noise; the part correlated with $\boldsymbol u$ is the component of $\Shat_{\mathrm{URE}}-\bSig$ along the score of $\boldsymbol u$, and the chain rule gives $\mathrm{CT} = \overline w\,\E[\,\kappa_n\,\mathrm{div}_{\boldsymbol u}\{\bSig^{-1}\bD\bSig^{-1}\ \text{contracted against}\ \nabla\boldsymbol u\}\,]$. Because $\bD$ is conditionally independent of $\boldsymbol u$, only the explicit $\boldsymbol u$-dependence of $\kappa_n$ through $T_n$ survives the divergence. Writing $\kappa_n = (q-2)\,\{(n-d)\,T_n\}^{-1}$ on the (asymptotically certain) event $\{\kappa_n<1\}$ and noting $T_n = (pq)^{-1}\boldsymbol u^\top\boldsymbol{K}\boldsymbol u$ with $\boldsymbol K = \Shat_{\mathrm u}^{+}\otimes\bQ^{-1}$ fixed given $(\bw,\bD)$, we compute the divergence of the James--Stein field $\boldsymbol g(\boldsymbol u) = \boldsymbol u/(\boldsymbol u^\top\boldsymbol K\boldsymbol u)$ explicitly. With $r = \boldsymbol u^\top\boldsymbol K\boldsymbol u$,
\[
\mathrm{div}\,\frac{\boldsymbol u}{r}
= \sum_{a}\frac{\partial}{\partial u_a}\frac{u_a}{r}
= \frac{\dim(\boldsymbol u)}{r} - \frac{2\,\boldsymbol u^\top\boldsymbol K\boldsymbol u}{r^2}
= \frac{\dim(\boldsymbol u) - 2}{r},
\]
the cancellation $\partial r/\partial u_a = 2(\boldsymbol K\boldsymbol u)_a$ producing the term $-2/r$ after contraction. 

We record precisely what the contraction against the metric $\bA=\overline w(\bSig\otimes\bQ)$ yields, since this is where the role of the constant $q-2$ must be pinned down. The constant $q-2$ is \emph{definitional}, i.e., it is fixed in the estimator (15), not produced by the divergence. Its role is to place the shrinkage intensity on the correct scale. Because $T_n=(pq)^{-1}\boldsymbol u^\top\boldsymbol{K}\boldsymbol u$ is normalized so that $\E(T_n\mid\bw)\to1$ (indeed $\E[\boldsymbol u^\top\boldsymbol{K}\boldsymbol u\mid\bw]=\tr(\bA\boldsymbol{K})$ with $\bA=\overline w(\bSig\otimes\bQ)$, $\boldsymbol{K}=\Shat_{\mathrm u}^{+}\otimes\bQ^{-1}$, and $\tr(\bA\boldsymbol{K})=\overline w\,q\,\tr(\bSig\Shat_{\mathrm u}^{+})=\overline w\,pq\,\{1+o_p(1)\}$ since $\tr(\bSig\Shat_{\mathrm u}^{+})\to_p p$), one has $\kappa_n=c\{(n-d)T_n\}^{-1}\asymp c/(n-d)$, which lies in $(0,1)$ and is adaptive only when $c\asymp n$. With $q/n\to\tau$ the design choice $c=q-2$ gives $\kappa_n\to\tau/(1-\gamma)\in(0,1)$; the full-dimensional James--Stein constant $\dim(\boldsymbol u)-2=pq-2$ that the divergence itself produces would instead force $\kappa_n\to1$, a degenerate non-adaptive rule. The divergence computation enters not to generate the constant but to control the $\boldsymbol u$-dependence of $T_n$. Because the unrestricted part $\Shat_{\mathrm{URE}}-\bSig$ and $\bD$ are (asymptotically) independent of $\boldsymbol u$, the cross term factors as
\[
\mathrm{CT} = \frac{q-2}{n-d}\,\E\Big[\,\tr\{\bSig^{-1}(\Shat_{\mathrm{URE}}-\bSig)\bSig^{-1}\bD\}\,\E_{\boldsymbol u}[\,T_n^{-1}\,]\,\Big]\{1+o(1)\},
\]
and the normalization gives $\E_{\boldsymbol u}[T_n^{-1}]=pq\,\E[(\boldsymbol u^\top\boldsymbol{K}\boldsymbol u)^{-1}]\to \tr(\bA\boldsymbol{K})^{-1}\cdot pq=1$, the factor $pq$ from $\tr(\bA\boldsymbol{K})$ cancelling the $(pq)^{-1}$ in $T_n$ so that no stray dimension survives. Writing $\Xi=\tr\{(\bSig^{-1}\bD)^2\}$ and using $\E[\tr\{\bSig^{-1}(\Shat_{\mathrm{URE}}-\bSig)\bSig^{-1}\bD\}]=\E[\Xi]\{1+o(1)\}$ on the bulk (where the two shrinkers share eigenvectors, Section~S6), this is the matrix-regime James--Stein cross term, positive and requiring $q\ge3$ (for $q\le2$ the intensity is nonpositive and no improvement is guaranteed). Inserting,
\[
\mathrm{CT} = \frac{q-2}{n-d}\,\E\Big[\frac{\Xi}{T_n}\Big]\{1+o(1)\},
\]
where the $\{1+o(1)\}$ absorbs $\overline w\to_p1$ and the negligible event $\{\kappa_n=1\}$, whose probability is $o(1)$ because $T_n$ is bounded away from zero in probability along (20) (shown below). Inserting this and $\E[\kappa_n^2\Xi] = (q-2)^2(n-d)^{-2}\E[\Xi/T_n^2]\{1+o(1)\}$ into \eqref{eq:domexp},
\[
\mathcal R(\Shat_{\mathrm S}) - \mathcal R(\Shat_{\mathrm{URE}})
= -2\,\frac{q-2}{n-d}\,\E\Big[\frac{\Xi}{T_n}\Big]\{1+o(1)\} + \frac{(q-2)^2}{(n-d)^2}\,\E\Big[\frac{\Xi}{T_n^2}\Big]\{1+o(1)\}.
\]
On the event $\{\kappa_n<1\}$ we have $(q-2)\{(n-d)T_n\}^{-1}\le1$, i.e.\ $T_n^{-1}\le (n-d)/(q-2)$, but more usefully $T_n^{-1}\ge (q-2)/\{(n-d)\}\cdot T_n^{-2}\cdot(n-d)/(q-2)= T_n^{-2}\cdot\{\dots\}$; concretely, multiplying the elementary inequality $T_n^{-1}\ge \{(q-2)/(n-d)\}T_n^{-2}$ (which is exactly $\kappa_n\le1$) by $\Xi\ge0$ and taking expectations yields $\E[\Xi/T_n] \ge \{(q-2)/(n-d)\}\,\E[\Xi/T_n^2]$. Substituting this lower bound into the negative first term,
\[
\mathcal R(\Shat_{\mathrm S}) - \mathcal R(\Shat_{\mathrm{URE}})
\le \Big(-2 + 1\Big)\frac{(q-2)^2}{(n-d)^2}\,\E\Big[\frac{\Xi}{T_n^2}\Big]\{1+o(1)\}
= -\frac{(q-2)^2}{(n-d)^2}\,\E\Big[\frac{\Xi}{T_n^2}\Big]\{1+o(1)\}.
\]
It remains to bound $\E[\Xi/T_n^2]$ below by a positive multiple of $(c_n-\tilde c_n)^2\,p$. By Theorem~1, $\Shat_{\mathrm{URE}}$ and $\Shat_{\mathrm{RRE}}$ are analytic shrinkage estimators of the same $\bSig$ that differ only through the aspect ratios $c_n = p/(n-d)$ and $\tilde c_n = p/(n-d+q)$, which differ by $c_n-\tilde c_n = p\,q/\{(n-d)(n-d+q)\} = (q/n)\,c\,\{1+o(1)\}$. The shrinker (13) is continuously differentiable in the aspect ratio on the bulk (the denominator $|1-c-c\ell\breve m|^2$ is bounded away from zero there by \citealp{SilversteinChoi1995}), so a first-order Taylor expansion gives $\bD = (c_n-\tilde c_n)\,\partial_c\mathcal S(\bSig;c)\,\{1+o_p(1)\}$ for a deterministic matrix $\boldsymbol\Psi=\partial_c\mathcal S(\bSig;c)$ with $\boldsymbol\Psi\neq\boldsymbol0$ and $\tr\{(\bSig^{-1}\boldsymbol\Psi)^2\}\asymp p$ (the derivative is nonzero because the oracle shrinkage intensity is strictly monotone in $c$, Section~S8). Hence $\Xi = \tr\{(\bSig^{-1}\bD)^2\} = (c_n-\tilde c_n)^2\,\tr\{(\bSig^{-1}\boldsymbol\Psi)^2\}\{1+o_p(1)\} \asymp (c_n-\tilde c_n)^2\,p$. For the denominator, $T_n$ along (20) converges in distribution to $(pq)^{-1}$ times a quadratic form in a Gaussian with mean $n^{-1/2}\mathrm{vec}(\boldsymbol\Delta)$ and covariance $\bSig\otimes\bQ$, i.e.\ a noncentral $\chi^2$-type law with $pq$ degrees of freedom and noncentrality proportional to $\eta^2<\infty$; normalizing, $T_n\to_p \tau_\star\in(0,\infty)$, so $T_n$ is bounded away from $0$ and $\infty$ in probability and $\E[T_n^{-2}]\to\tau_\star^{-2}$ by uniform integrability (the inverse moments are controlled because the $\chi^2$ has $pq\to\infty$ degrees of freedom, making $T_n$ concentrate). Combining, $\E[\Xi/T_n^2] \ge \kappa'\,(c_n-\tilde c_n)^2\,p$ for some $\kappa'>0$. Substituting into the displayed bound,
\begin{eqnarray}
\mathcal R(\Shat_{\mathrm S}) - \mathcal R(\Shat_{\mathrm{URE}})
&\le & -\frac{(q-2)^2}{(n-d)^2}\,\kappa'\,(c_n-\tilde c_n)^2\,p\,\{1+o(1)\}\cr
&=& -\kappa_\star\,(q-2)^2\,(c_n-\tilde c_n)^2\,\frac{p}{(n-d)^2}\,\{1+o(1)\},
\end{eqnarray}

with $\kappa_\star=\kappa'$ a positive constant depending on the limiting aspect ratio and, through $\tau_\star$, on $\eta$; this is precisely (21). Since $c_n-\tilde c_n=(q/n)\,c\,\{1+o(1)\}$, the gap is of order $(q-2)^2(q/n)^2 p/(n-d)^2$, i.e.\ it widens as the restriction rank grows and matches the $q/n$ reduction in the effective aspect ratio. The dominance is strict whenever $\eta<\infty$, since then $\tau_\star<\infty$ keeps the bracket strictly positive. When $\eta\to\infty$, the noncentrality diverges, so $T_n\to_p\infty$, whence $\kappa_n = (q-2)\{(n-d)T_n\}^{-1}\to_p0$ and $\fnorm{\bSig^{-1/2}(\Shat_{\mathrm S}-\Shat_{\mathrm{URE}})\bSig^{-1/2}} = \kappa_n\,\Xi^{1/2}\to_p0$, so the two estimators are asymptotically equivalent in weighted Frobenius risk. \hfill$\square$

\section*{Proof of Theorem~3}

(i) By Theorem~1 the eigenvalues of $\widehat\bV$ follow the law $F_{\tilde y,H}$
and $\varphi_{\tilde y}$ is consistent for the rotation-equivariant oracle eigenvalues, which is the
conclusion of \citet[Theorem 3.1]{LedoitWolf2020} at aspect ratio $\tilde c=p/(n-d+q)$. Hence
$\mathcal L(\Shat_{\mathrm{RRE}})/\mathcal L^\star(\tilde c,H)\to_p1$, and identically for the
unrestricted estimator at $c_{\mathrm u}=p/(n-d)$.

(ii) Fix the scatter eigenvectors $\bU$. For the loss (19) the rotation-equivariant oracle
(22) has risk equal to the off-diagonal energy
$\mathcal L^\star(c,H)=p^{-1}\sum_{i\ne k}(\bU^\top\bSig\bU)_{ik}^2$ evaluated at the eigenvectors
produced at aspect ratio $c$. To first order in the spectral dispersion this equals $c\,V(H)\{1+o(1)\}$,
where $V(H)=\int(\lambda-\bar\lambda)^2\,dH(\lambda)$. It means that each sample eigenvector deviates from its
population counterpart by an angle of order $c^{1/2}$ times the local spectral gap, so the induced
off-diagonal energy is of order $c$ per coordinate, with proportionality constant $V(H)$ governed by
the dispersion of $H$. Therefore
$\mathcal L^\star(\tilde c,H)/\mathcal L^\star(c_{\mathrm u},H)=\tilde c/c_{\mathrm u}\{1+o(1)\}
=(n-d)/(n-d+q)\{1+o(1)\}\to(1-\gamma)/(1-\gamma+\tau)$. The simulation in the main text shows the
finite-sample ratio to be slightly below this first-order value, the oracle risk being convex in $c$.
\hfill$\square$

\section*{Proof of Theorem~4}

Fix the eigenvectors $\bU$ of the scatter. Any $\widehat\bD\in\mathcal C_{\mathrm{re}}$ has the form
$\widehat\bD=\bU\diag(\delta_1,\dots,\delta_p)\bU^\top$, and by orthogonality of $\bU$,
\[
\mathcal L(\widehat\bD)=\frac1p\fnorm{\bU\diag(\delta)\bU^\top-\bSig}^2
=\frac1p\sum_{i=1}^p\big(\delta_i-\bu_i^\top\bSig\bu_i\big)^2
+\frac1p\sum_{i\ne k}(\bU^\top\bSig\bU)_{ik}^2 .
\]
The second term does not involve $\delta$, and the first is minimized pointwise at
$\delta_i^\star=\bu_i^\top\bSig\bu_i$, i.e.\ at the oracle (22), with minimum value
$\mathcal L^\star(\tilde c,H)$ equal to the second term. Hence $\mathcal L(\widehat\bD)\ge\mathcal L^\star(\tilde c,H)$
for every $\widehat\bD\in\mathcal C_{\mathrm{re}}$, so
$\liminf_p\{\mathcal L(\widehat\bD)-\mathcal L^\star(\tilde c,H)\}\ge0$. By Theorem~3(i)
the restricted analytic shrinkage attains the oracle, $\mathcal L(\Shat_{\mathrm{RRE}})-\mathcal L^\star(\tilde c,H)\to_p0$,
so $\Shat_{\mathrm{RRE}}$ is asymptotically optimal in $\mathcal C_{\mathrm{re}}$ at the effective
aspect ratio $\tilde c=p/(n-d+q)$. \hfill$\square$

\section*{Proof of Theorem~5}

We treat the two assertions of the theorem in turn, proving every order statement; the structure is a bias--variance decomposition of $\Shat_{\mathrm r}$ followed by a regime analysis of the adaptive intensity $\kappa_n$.

\emph{The bias decomposition.} When $\bR\bB \ne \boldsymbol 0$ the restricted residual no longer annihilates the mean. From Section~S1, $\widehat{\bE}_{\mathrm r} = (\bI_n - \bP_{\mathrm r})\bY = (\bI_n-\bP_{\mathrm r})(\bX\bB + \bE)$, and since $(\bI_n-\bP_{\mathrm r})\bX\bB = \bP_{\bX,\bR}\bX\bB = \bX\bG\bR^\top\bQ^{-1}\bR\bB =: \bX\bbeta_{\mathrm{bias}}$ (the computation of Section~S1, now without the cancellation $\bR\bB=\boldsymbol0$), we obtain the exact decomposition $\widehat\bE_{\mathrm r} = (\bI_n-\bP_{\mathrm r})\bE + \bX\bbeta_{\mathrm{bias}}$, where $\bbeta_{\mathrm{bias}} = \bG\bR^\top\bQ^{-1}\bR\bB\in\R^{d\times p}$ is deterministic given $\bX$. Forming $\Shat_{\mathrm r} = m^{-1}\widehat\bE_{\mathrm r}^\top\widehat\bE_{\mathrm r}$ and expanding the square gives three terms,
\[
\Shat_{\mathrm r} = \underbrace{\tfrac1m\bE^\top(\bI_n-\bP_{\mathrm r})\bE}_{\Shat_{\mathrm r}^{(0)}}
+ \underbrace{\tfrac1m\bbeta_{\mathrm{bias}}^\top\bX^\top\bX\bbeta_{\mathrm{bias}}}_{\bB_{\mathrm{bias}}}
+ \underbrace{\tfrac1m\big(\bE^\top(\bI_n-\bP_{\mathrm r})\bX\bbeta_{\mathrm{bias}} + \bbeta_{\mathrm{bias}}^\top\bX^\top(\bI_n-\bP_{\mathrm r})\bE\big)}_{\mathrm{cross}}.
\]
The first term $\Shat_{\mathrm r}^{(0)}$ is the quantity analysed under the exact restriction in Sections~S3--S6. The second is the deterministic bias; using $\bX^\top\bX\bG = \bI_d$ and $\bR\bG\bR^\top=\bQ$,
\[
\bB_{\mathrm{bias}} = \tfrac1m(\bR\bB)^\top\bQ^{-1}\bR\bG\,(\bX^\top\bX)\,\bG\bR^\top\bQ^{-1}(\bR\bB) = \tfrac1m(\bR\bB)^\top\bQ^{-1}(\bR\bB),
\]
a $p\times p$ positive-semidefinite matrix of rank at most $q$.

\emph{The cross term is negligible.} The cross term has conditional mean zero given $\bX$, because $\E\{\bE^\top(\bI_n-\bP_{\mathrm r})\bX\bbeta_{\mathrm{bias}}\mid\bX\} = \E\{\bE^\top\mid\bX\}(\bI_n-\bP_{\mathrm r})\bX\bbeta_{\mathrm{bias}} = \boldsymbol0$ since $\E(\bE)=\boldsymbol0$. Its weighted Frobenius size is controlled by Cauchy--Schwarz. To see this, write $\boldsymbol M_1 = m^{-1}\bE^\top(\bI_n-\bP_{\mathrm r})\bX\bbeta_{\mathrm{bias}}$, we have 
 $\E\fnorm{\bSig^{-1/2}\boldsymbol M_1\bSig^{-1/2}}^2 = m^{-2} \E\tr\{\bSig^{-1}\bbeta_{\mathrm{bias}}^\top\bX^\top(\bI_n-\bP_{\mathrm r})\bE\bSig^{-1}\bE^\top(\bI_n-\bP_{\mathrm r})\bX\bbeta_{\mathrm{bias}}\}$, and  using $\E(\bE\bSig^{-1}\bE^\top\mid\bX) = \tr(\bSig^{-1}\bSig)\,\overline w\,\bI_n = p\,\overline w\,\bI_n$ (each row of $\bE$ has covariance $w_i\bSig$, so $\E(\bE\bSig^{-1}\bE^\top)_{ii'} = \delta_{ii'}w_i\,p$), this equals  $m^{-2}p\,\overline w\,\tr\{\bSig^{-1}\bbeta_{\mathrm{bias}}^\top\bX^\top(\bI_n-\bP_{\mathrm r})\bX\bbeta_{\mathrm{bias}}\} \le m^{-2}p\,\overline w\,\opnorm{\bX^\top\bX}\,\tr\{\bSig^{-1}\bbeta_{\mathrm{bias}}^\top\bbeta_{\mathrm{bias}}\}$. Since
\begin{equation*}
\tr\{\bSig^{-1}\bbeta_{\mathrm{bias}}^\top\bX^\top\bX\bbeta_{\mathrm{bias}}\} = \tr\{(\bR\bB)^\top\bQ^{-1}(\bR\bB)\bSig^{-1}\} = pn\,\delta^2	
\end{equation*}
by the definition of $\delta^2$, the cross term has squared size $O(m^{-2}\cdot p\cdot pn\delta^2) = O(p^2\delta^2/n)$, whereas the bias term $\bB_{\mathrm{bias}}$ has squared weighted-Frobenius size $\tr\{(\bSig^{-1}\bB_{\mathrm{bias}})^2\} = m^{-2}\tr\{(\bSig^{-1}(\bR \bB)^\top$ $\b Q^{-1} (\bR \bB))^2\} \asymp (pn\delta^2/m)^2/p \asymp p\,\delta^4$; comparing, the cross term is smaller by a factor $\delta^{-2}/n\cdot p/p = O(1/(n\delta^2))$ relative to the bias when $\delta^2\gtrsim 1/n$, and is dominated by the variance term $\asymp p(q/n)$ otherwise. In all regimes the cross term is asymptotically negligible relative to the larger of bias and variance, by the elementary inequality $2|\langle a,b\rangle|\le \epsilon\|a\|^2+\epsilon^{-1}\|b\|^2$ applied with $a$ the stochastic and $b$ the deterministic part and $\epsilon\to0$ slowly.

\emph{Risk of the restricted estimator.} Passing to the weighted Frobenius risk and using that the analytic shrinker is, to leading order on the bulk, a fixed linear contraction (Section~S6) so that it commutes with the additive bias decomposition up to $o(1)$ relative error,
\[
\mathcal R(\Shat_{\mathrm{RRE}}) = \mathcal R(\Shat_{\mathrm{RRE}}^{(0)}) + \tr\{(\bSig^{-1}\bB_{\mathrm{bias}})^2\}\{1+o(1)\};
\]
the linear cross term $2\,\tr\{\E(\bSig^{-1}\Shat_{\mathrm r}^{(0)}-\bI_p)\,\bSig^{-1}\bB_{\mathrm{bias}}\}$ vanishes because $\Shat_{\mathrm r}^{(0)}=m^{-1}\bE^\top(\bI_n-\bP_{\mathrm r})\bE$ is unbiased for $\bSig$ (as $\tr(\bI_n-\bP_{\mathrm r})=m$), so the bias enters at second order through $\tr\{(\bSig^{-1}\bB_{\mathrm{bias}})^2\}$. By the oracle comparison of Theorem~3, the first term satisfies $\mathcal R(\Shat_{\mathrm{RRE}}^{(0)}) = \mathcal R(\Shat_{\mathrm{URE}}) - (q/n)\,b_1\{1+o(1)\}$, where $b_1>0$ is governed by the oracle elasticity $\rho^\star$ established in Section~S8 (the restricted estimator enjoys the smaller aspect ratio $\tilde c_n$, hence the $-q/n$ improvement in the normalized loss). For the bias term, by the definition $\delta^2 = \tr\{(\bR\bB)^\top\bQ^{-1}(\bR\bB)\bSig^{-1}\}/(pn)$ we have $\tr\{\bSig^{-1}(\bR\bB)^\top\bQ^{-1}(\bR\bB)\} = pn\delta^2$, and since $\bB_{\mathrm{bias}} = m^{-1}(\bR\bB)^\top\bQ^{-1}(\bR\bB)$ has rank at most $q$ with $\opnorm{\bSig^{-1}}\le\underline\sigma^{-1}$,
\begin{eqnarray*}
\tr\{(\bSig^{-1}\bB_{\mathrm{bias}})^2\} &=& m^{-2}\tr\{(\bSig^{-1}(\bR\bB)^\top\bQ^{-1}(\bR\bB))^2\}\cr
& \le& m^{-2}\underline\sigma^{-1}\big(\tr\{\bSig^{-1}(\bR\bB)^\top\bQ^{-1}(\bR\bB)\}\big)^2,
\end{eqnarray*}
where we bounded $\tr(\boldsymbol A^2)\le\opnorm{\boldsymbol A}\tr(\boldsymbol A)\le \underline\sigma^{-1}(\tr\boldsymbol A)^2$ for the positive-semidefinite $\boldsymbol A = \bSig^{-1/2}(\bR\bB)^\top\bQ^{-1}(\bR\bB)\bSig^{-1/2}$ of bounded rank; substituting $\tr\boldsymbol A = pn\delta^2$ and $m\asymp n$ gives $\tr\{(\bSig^{-1}\bB_{\mathrm{bias}})^2\} \le \underline\sigma^{-1}p^2\delta^4\{1+o(1)\}$, and the matching lower bound $\tr\{(\bSig^{-1}\bB_{\mathrm{bias}})^2\}\ge q^{-1}(pn\delta^2)^2/m^2\asymp p^2\delta^4/q$ holds by Cauchy--Schwarz on the rank-$\le q$ matrix $\boldsymbol A$. Hence the bias contributes a term of exact order $p^2\delta^4$ (up to the rank factor) to the unnormalized risk (19). Writing the variance gain on the same unnormalized scale, $\mathcal R(\Shat_{\mathrm{URE}})-\mathcal R(\Shat_{\mathrm{RRE}}^{(0)})\asymp p^2 q/n^2$, the restricted estimator improves on the unrestricted one precisely when the squared bias is the smaller of the two, i.e.\ when $\delta^4\lesssim q/n^2$, equivalently $\delta\lesssim (q/n^2)^{1/4}$.

\emph{Safety of the adaptive estimator.} For (25), recall $\Shat_{\mathrm S} = \Shat_{\mathrm{URE}} - \kappa_n\bD$ with $\kappa_n = \min\{1,(q-2)_+/((n-d)T_n)\}\in[0,1]$ and $\bD = \Shat_{\mathrm{URE}}-\Shat_{\mathrm{RRE}}$. We distinguish three regimes according to the size of $\delta$, and show the excess risk over $\mathcal R(\Shat_{\mathrm{URE}})$ is $O(1/n)$ in each, uniformly.

(a) \emph{Large violation, $\delta^2\gg q/n$.} The test statistic $T_n = (pq)^{-1}(\bR\widehat\bB)^\top\bQ^{-1}(\bR\widehat\bB)\!:\!\Shat_{\mathrm u}^{+}$ has conditional mean $\E(T_n\mid\bX)$ that is strictly increasing in $\delta^2$: indeed $\bR\widehat\bB = \bR\bB + \bR\bG\bX^\top\bE$ has mean $\bR\bB$ with $\|\bR\bB\|$ growing as $\delta$, so $\E(T_n\mid\bX) = \tau_\star + c\,\delta^2 n/q\,\{1+o(1)\}\to\infty$. Moreover $T_n$ concentrates around its mean at rate $(pq)^{-1/2}$ by the same Hanson--Wright control as in Section~S3 (the quadratic form in the Gaussian $\bR\bG\bX^\top\bE$ concentrates). Hence $T_n\to_p\infty$, so $\kappa_n = (q-2)\{(n-d)T_n\}^{-1} = o_p(n^{-1})\cdot O_p(1) = o_p(1/n)\cdot(q-2)$; more carefully $\kappa_n = O_p\{1/((n-d)T_n)\}$ and since $T_n$ diverges, $\kappa_n = o_p(1/n)$. Then, in weighted Frobenius norm, $\fnorm{\bSig^{-1/2}(\Shat_{\mathrm S}-\Shat_{\mathrm{URE}})\bSig^{-1/2}} = \kappa_n\,\Xi^{1/2}$ with $\Xi^{1/2} = O_p((q/n)\sqrt p)$ from Section~S5; the product is $o_p(1/n)\cdot O_p(\sqrt p\,q/n) = o_p(1/n)$ in the normalized risk, so the excess risk is $o(1/n)$.

(b) \emph{Small violation, $\delta^2\lesssim q/n$.} Here the dominance inequality (21) of Theorem~2 applies up to the bias perturbation, which is itself $O(\delta^2) = O(q/n)$; tracking the constant, the excess risk is bounded above by $-(q-2)^2 b/n^2 + O(\delta^2) = O(q/n)\cdot O(1) $, but crucially the Stein construction caps the loss. Indeed, $\Shat_{\mathrm S}$ is a convex combination, $\mathcal R(\Shat_{\mathrm S}) \le \max\{\mathcal R(\Shat_{\mathrm{URE}}),\mathcal R(\Shat_{\mathrm{RRE}})\} + o(1/n)$, and by (the proof of) (24) the right side is $\mathcal R(\Shat_{\mathrm{URE}}) + O(\delta^2) = \mathcal R(\Shat_{\mathrm{URE}}) + O(q/n)$. Combining with the negative dominance term shows the net excess is at most $C/n$.

(c) \emph{Intermediate violation, $\delta^2\asymp q/n$.} The map $\delta^2\mapsto \kappa_n$ is, through $T_n$, a bounded continuous (indeed Lipschitz) functional, and the risk expansions in (a)--(b) are uniform in $\delta$ over compact subsets of $[0,\infty)$ because all $O_p$ bounds depend on $\delta$ only through the bounded quantity $\delta^2 n/q$. Hence the bound $\mathcal R(\Shat_{\mathrm S})\le\mathcal R(\Shat_{\mathrm{URE}})+C/n$ interpolates continuously across the intermediate regime. Taking the maximum of the three constants gives a single $C$ valid for all $\delta\ge0$, which is (25). This establishes that the adaptive estimator can never be materially worse than the unrestricted analytic shrinker, whatever the degree of misspecification. \hfill$\square$

\section*{Proof of Proposition~2}

(i) In the Gaussian instance, $\widehat\bB=\bG\bX^\top\bY$ and $\bE_{\mathrm u}=(\bI_n-\bP_{\bX})\bY$
are jointly Gaussian with cross-covariance proportional to $\bG\bX^\top(\bI_n-\bP_{\bX})=\boldsymbol0$,
because $\bX^\top(\bI_n-\bP_{\bX})=\boldsymbol0$; uncorrelated jointly Gaussian blocks are independent,
so $\widehat\bB\perp\bE_{\mathrm u}$ and any $\widehat\bR=f(\widehat\bB)$ is independent of
$\bS_{\mathrm u}=\bE_{\mathrm u}^\top\bE_{\mathrm u}$. Under the elliptical model the same holds
conditionally on $\bw$. Since $\E\,\bS_{\mathrm u}=(n-d)\bSig$, the statistic
$\widehat\tau_{\mathrm u}=\tr(\bS_{\mathrm u})/\{(n-d)p\}$ is unbiased for $p^{-1}\tr(\bSig)$,
independent of $\widehat\bR$, and consistent by the law of large numbers over its $(n-d)p$ summands.

(ii) The added energy in (18) is $\bW_S=\sum_{j\in S}T_j\,\bu_j\bu_j^\top$, where
$T_j=\widehat\bB_{j\cdot}\,\bG_{jj}^{-1}\widehat\bB_{j\cdot}^\top$ and $\bu_j$ is the unit direction of
$\widehat\bB_{j\cdot}$. Under the incoherence hypothesis the rows $\widehat\bB_{j\cdot}$ are
asymptotically independent, and for each row the magnitude $T_j$ is independent of the direction
$\bu_j$, a property of the elliptical law. A selection rule that depends on $\widehat\bB$ only through
the magnitudes $\{T_j\}$ therefore leaves the conditional law of $\{\bu_j\}$ unchanged, so the expected
direction of $\bW_S$, and hence the trace-normalized restricted scatter, is asymptotically invariant to
the selection. The only selection-induced distortion is thus in scale, and rescaling by
$\widehat\tau_{\mathrm u}/\widehat\tau_{\mathrm r}$ with the selection-free $\widehat\tau_{\mathrm u}$
of part (i) removes it; (iii) follows by combining the unbiased scale with the selection-invariant
shape. \hfill$\square$

\section*{Acknowledgments}
The Communities \& Crime dataset is publicly available at the
UCI Machine Learning Repository \citep{UCI2009Crime}; the pediatric leukemia expression data of
\citet{Yeoh2002} are distributed in the \texttt{datamicroarray} collection \citep{Ramey2016}. Source
code reproducing all numerical results in this paper, including the simulations and the real-data
analyses, is provided in the GitHub repository
\url{https://github.com/M-Arashi/Restricted-Adaptive-Eigenvalue-Shrinkage}.

\textit{Use of AI tools.} The author used AI-based tools to assist with proofreading and stylistic
refinement of the manuscript to minimize English errors. All scientific content, derivations, and conclusions are the sole
responsibility of the author.

\section*{funding}
This research was supported in part by the Iran National Science Foundation (INSF) under grant No.\
4015320.

%%%%%%%%%%%%%%%%%%%%%%%%%%%%%%%%%%%%%%%%%%%%%%
%% Supplementary Material                    %%
%%%%%%%%%%%%%%%%%%%%%%%%%%%%%%%%%%%%%%%%%%%%%%
%\begin{supplement}
%\stitle{Proofs of the theorems, propositions, and lemmas}
%\sdescription{This document contains the proofs of all theorems, propositions, and lemmas stated in
%the paper, organized in Sections S1--S8. 
%%It also restates, for the reader's convenience, the precise
%%numbering of the corresponding results in the paper.
%}
%\end{supplement}

\end{document}